\documentclass[12pt,preprint]{emulateapj}

\usepackage{color}
\usepackage{ascmac}

\slugcomment{The Astrophysical Journal in press}

\shorttitle{Properties of Interstellar Dust}
\shortauthors{Nozawa \& Fukugita}

\begin{document}

\title{PROPERTIES OF DUST GRAINS PROBED WITH EXTINCTION CURVES}

\author{
Takaya Nozawa\altaffilmark{1} and
Masataka Fukugita\altaffilmark{1,2,3}
}

\altaffiltext{1}{Kavli Institute for the Physics and Mathematics of the
Universe, University of Tokyo, Kashiwa, 277-8583, Japan} 
\altaffiltext{2}{Institute for Cosmic Ray Research, University of Tokyo,
Kashiwa, 277-8582, Japan}
\altaffiltext{3}{Institute for Advanced Study, Princeton, NJ 08540, USA}

%%%%%%%%%%%%%%%%%%%%%%%%%%%%%%%%%%%%%%%%%%%%%%%%%%%%%%%%%%%%%%%%%%%%%%
\begin{abstract}

Modern data of the extinction curve from the ultraviolet to the near
infrared are revisited to study properties of dust grains in the Milky
Way (MW) and the Small Magellanic Cloud (SMC).  
We confirm that the graphite-silicate mixture of grains yields the 
observed extinction curve with the simple power-law distribution of 
the grain size but with a cutoff at some maximal size: 
the parameters are tightly constrained to be $q = 3.5 \pm 0.2$ for 
the size distribution $a^{-q}$ and the maximum radius 
$a_{\rm max} =0.24 \pm 0.05$ $\mu$m, for both MW and SMC.  
The abundance of grains, and hence the elemental abundance, is 
constrained from the reddening versus hydrogen column density, 
$E(B-V)/N_{\rm H}$.  
If we take the solar elemental abundance as the standard for the MW, 
$>$56 \% of carbon should be in graphite dust, while it is $<$40 \% 
in the SMC using its available abundance estimate.  
This disparity and the relative abundance of C to Si explain the 
difference of the two curves.  
We find that 50--60 \% of carbon may not necessarily be in graphite 
but in the amorphous or glassy phase.  
Iron may also be in the metallic phase or up to $\sim$80 \% in 
magnetite rather than in silicates, so that the Mg/Fe ratio in 
astronomical olivine is arbitrary.  
With these substitutions the parameters of the grain size remain 
unchanged.  
The mass density of dust grains relative to hydrogen is 
$\rho_{\rm dust} / \rho_{\rm H} = 1 / (120 {+10 \atop -16})$ for the 
MW and $1 / (760 {+70 \atop  -90})$ for the SMC under the elemental 
abundance constraints.  
We underline the importance of the wavelength-dependence of the
extinction curve in the near infrared in constructing the dust model:
if $A_\lambda \propto \lambda^{-\gamma}$ with $\gamma \simeq 1.6$, 
the power-law grain-size model fails, whereas it works if 
$\gamma \simeq$ 1.8--2.0.

\end{abstract}
%%%%%%%%%%%%%%%%%%%%%%%%%%%%%%%%%%%%%%%%%%%%%%%%%%%%%%%%%%%%%%%%%%%%%%


%%%%%%%%%%%%%%%%%%%%%%%%%%%%%%%%%%%%%%%%%%%%%%%%%%%%%%%%%%%%%%%%%%%%%%
\section{Introduction}
%%%%%%%%%%%%%%%%%%%%%%%%%%%%%%%%%%%%%%%%%%%%%%%%%%%%%%%%%%%%%%%%%%%%%%

Submicron-size grains cause attenuation of light from ultraviolet (UV)
to near-infrared (NIR) wavelengths. 
Larger grains may form small astronomical objects.  
Thirty years ago, Mathis, Rumpl, \& Nordsieck (1977, hereinafter MRN) 
showed that dust composed of a mixture of silicate and carbonaceous 
grains accounts for the extinction curve from the UV to optical 
wavelengths, from which they derived the size distribution of grains 
for $a=0.005$ to 0.25 $\mu$m.  
Only with 9 bins of the histogram, they advocated that the size 
distribution is consistent with the power-law $a^{-q}$ with the index 
$q$ from 3.3 to 3.6.\footnote{For a reference we attempt to fit the 
  size distribution derived by MRN (5 data points for silicate, and 6 
  for graphite) by the power-law: it results in $q = 3.8 \pm 0.7$ for 
  graphite and $q = 3.5 \pm 0.6$ for silicate, where the error stands 
  for the dispersion.}  
This dust model is consistent with the fact that Mg, Si, and Fe are 
highly depleted in diffuse interstellar matter.  
These three elements are also major constituents of astronomical 
silicate with the corroborating fact that the abundance of dust needed 
to account for the observed extinction is on the order of magnitude of
their cosmic abundance, and hence the three elements could dominantly
be locked in dust.

Independently, the argument appeared that small astronomical bodies in
frequent collisions would obey a power-law distribution in size
(Dohnanyi 1969; Hellyer 1970; Biermann \& Harwit 1980; Dorschner
1982). 
It was argued that the power-law derived by collisional equilibrium has 
typically $q=3.5$ (Dohnanyi 1969; Pan \& Sari 2005), which agrees with 
the power favoured by interstellar dust from extinction studies.  
For larger sizes the distribution of small astronomical objects 
($\gtrsim$ 1--10 km) does not conflict with this power, while the 
available information is limited to derive it more accurately .

These arguments indicate that the size distribution of grains may be 
key to understanding the formation of grains and small astronomical
objects.  
The size distribution of grains has been studied in the literature 
(e.g., Draine \& Lee 1984, hereafter DL84; Kim, 
Martin, \& Hendry 1994; Weingartner \& Draine 2001, hereafter WD01; 
Clayton et al.\ 2003a; Zubko, Dwek, \& Arendt 2004).
These studies stress accurately reproducing the
extinction curve that is progressively more accurately measured and is
extended to the NIR region, to find the detailed size distribution of
grains.  
A significant variation has become apparent in the extinction curve 
depending on the line of sight 
(Mathis \& Cardelli 1992; Clayton et al.\ 2000).  
The variation typically amounts to $R_V =$ 2.2 to 5.5, assuming the 
one-parameter formula of Cardelli, Clayton, \& Mathis (1989,
hereafter CCM: their formula is referred to as the CCM curve), 
who showed that the family of extinction curves is well
described by a formula with the single free parameter, $R_V = A_V /
E(B - V)$, where $A_V$ is the extinction in the $V$ band and $E(B-V)$
is the reddening.  
It should be asked how the variation of the extinction curve 
translates to the properties of dust grains. 
Another important problem is determining other grain
constituents beyond 
silicate and graphite.

Further to the shape of the extinction curve, the amounts of  
extinction, usually represented with $E(B-V)$ per hydrogen
gives an important constraint on the abundance of dust grains.  
An important question is whether the grain parameters derived from the
extinction curve are consistent with those from the amount of extinction, 
and whether they are consistent with the elemental abundance.

It has been known that the extinction curve for the Small Magellanic 
Cloud (SMC) is markedly different from that for the Milky Way (MW) in 
that it lacks the feature at 2175 \AA~and it shows a significantly 
steeper rise to the far UV side beyond 2000 \AA.  
What would cause this difference is the problem to be asked.  
The SMC type extinction curve is also indicated for interstellar 
clouds such as Mg II absorbers (York et al.\ 2006); 
it is shown that the majority, typically $\sim$70 \%, of the cloud 
obeys the SMC type extinction law.

In this paper we revisit gross but generic features of dust
grains using the modern data of extinction curves from 1150 \AA~to
2.2 $\mu$m derived for the MW and, in the other extreme, the SMC, 
along with modern optical data of the relevant dielectric material.  
We examine whether some departure from power-law size distributions,
other than cutoffs in the size, is compelling in reproducing the
observed extinction curve.  
We are also interested in how dust in the SMC should differ from that 
in the MW in its properties. 
Furthermore, we would like to see whether the graphite-silicate model 
is unique and what is the possible range of dust to hydrogen mass ratio.  
We assume spherical grains and their size distribution obeying a 
power-law allowing for a truncation at some maximum size.  
The power-law distribution, at least, can be understood with simple
physics for the evolution of grains.  
We do not treat polarisation and infrared emission from dust since 
either requires knowledge other than the property of dust grains 
and so, in turn, requires extra assumptions that would introduce 
further uncertainties.

In Section 2, we review the data of extinction curves used in the
present study, and we note a problem in the average NIR extinction
curve.  
After defining the model of interstellar dust used in this paper in 
Section 3, we search, in Section 4, for the dust model that could 
reproduce the observation, present parameters of the model and ask 
various possibilities of the favourable composition of grain species 
from the extinction data for the MW. 
We carry out a similar analysis for SMC dust in Section 5.  
Section 6 gives a summary of this study with discussion.

%%%%%%%%%%%%%%%%%%%%%%%%%%%%%%%%%%
\begin{deluxetable}{llccc}
\tablewidth{0pt}
\tablecaption{Interstellar Extinction Data for the Milky Way}
\tablehead{
\multicolumn{2}{l}{Reference Wavelength} &
\multicolumn{1}{c}{$1/\lambda$} &
\multicolumn{2}{c}{($A_\lambda/A_V$)$_{\rm obs}$} \\
\multicolumn{1}{c}{($\mu$m)} &  &
\multicolumn{1}{c}{($\mu$m$^{-1}$)} &
\multicolumn{1}{c}{lower} &
\multicolumn{1}{c}{upper} 
}
\startdata
  & & & &  \\
0.125  & (far-UV rise) & 8.00 & 2.573 & 3.894 \\
0.16   & (far-UV dip)  & 6.25 & 2.167& 3.003 \\
0.2175 & (UV bump)     & 4.60 & 2.712 & 3.625 \\
0.36   & ({\it U} band)   & 2.78 & 1.449 & 1.62 \\
0.44   & ({\it B} band)   & 2.27 & 1.252 & 1.331 \\
0.55   & ({\it V} band)   &  1.82 & 1 & 1 \\
1.25   & ({\it J} band)   & 0.80 & 0.211 & 0.278 \\
1.65   & ({\it H} band)   & 0.61 & 0.127 & 0.166 \\
2.17   & ({\it K} band)   & 0.46 & 0.077 & 0.101 \\
\tableline
  & & & &  \\
\multicolumn{5}{c}{Ranges from the CCM formula for NIR} \\
\tableline
  & & &  & \\
1.25   & ({\it J} band)   & 0.80 & 0.267 & 0.299 \\
1.65   & ({\it H} band)   & 0.61 & 0.171 & 0.191\\
2.17   & ({\it K} band)   & 0.46 & 0.110 & 0.123\\
\enddata
%%%%%%%%
\tablecomments{
The allowed ranges of the interstellar extinction 
($A_\lambda/A_V$)$_{\rm obs}$ at the reference wavelengths 
constructed from the data of extinction curves of FM07.
The ranges of the NIR extinction from the CCM formula is for 
$R_V =$ 2.75--3.60.
In optical regions, the ranges virtually agree with the 1 $\sigma$ 
ranges of FM07.
}
\end{deluxetable}
%%%%%%%%%%%%%%%%%%%%%%%%%%%%%%%%%%

%%%%%%%%%%%%%%%%%%%%%%%%%%%%%%%%%%%%%%%%%%%%%%%%%%%%%%%%%%%%%%%%%%%%%%
\section{Extinction curves}
%%%%%%%%%%%%%%%%%%%%%%%%%%%%%%%%%%%%%%%%%%%%%%%%%%%%%%%%%%%%%%%%%%%%%%

%%%%%%%%%%%%%%%%%%%%%%%%%%%%%%%%%%%%%%%%%%%%%%%%%%%%%%%%%%%%%%%%%%%%%%
\subsection{Milky Way extinction}
%%%%%%%%%%%%%%%%%%%%%%%%%%%%%%%%%%%%%%%%%%%%%%%%%%%%%%%%%%%%%%%%%%%%%%

We take the extinction curves derived from UV to NIR for 328 stars in
Fitzpatrick \& Massa (2007, hereafter FM07) with the aid of stellar
atmosphere, replacing the traditional method using reddened-comparison
pairs of stars.  
The extinction curve ranges from 1150 \AA~to 2.2 $\mu$m.  
We are primarily interested in the global average of the extinction 
curves, so we consider a set of the curves at several specific 
reference wavelengths, at which we derive the allowed ranges
of extinction.  
We take 9 reference wavelengths in total with the {\it V} band to give 
the normalisation.  
We consider wavelengths corresponding to {\it UBJHK} and three 
wavelengths, $\lambda = 0.2175$ $\mu$m, 0.16 $\mu$m, and 0.125 $\mu$m 
in UV, which characterise the hump, the bottom in the UV region, and 
an arbitrary chosen wavelength in the rise of the Galactic extinction 
curve towards shorter wavelengths, respectively.  
The {\it UBVJHK} passbands are those FM07 used to derive the extinction 
curve in the optical and NIR regions from the observation (they have 
not used $R$ and $I$).  
We take the 1-$\sigma$ allowed ranges corresponding to 68 \% (224 
curves) of the 328 curves at each reference point.  
Our wavelength mesh is too coarse to study the 2175 \AA~feature; 
so it is considered separately.  
We do not treat diffuse interstellar bands which are not apparent in 
the FM07 extinction curve.  
Possible line features beyond the {\it K} band are not treated.

The 1-$\sigma$ ranges of $A_\lambda/A_V$ are shown in Table 1 and in
Figure 1: (a) for the entire wavelength range considered and (b) for
NIR in an expanded scale.\footnote{If we derive the extinction for 
  $R_C$ and $I_C$ from FM07 in the same manner as the other passbands, 
  the 1 $\sigma$ ranges are 0.722--0.778 and 0.492--0.587, respectively.  
  We note that these data are not directly constrained by observations.  
  In fact, direct observational determinations for the extinction for 
  $R_C$ and $I_C$ are scanty.
  Winkler (1997) estimated it assuming the fiducial colours for
  estimated type of stars. 
  His values overlap with those from FM07 at 1 $\sigma$ only marginally.  
  The extinction in the $I_C$ band discussed in Draine (2003a) extends 
  from $-1.5$ to +2.7 $\sigma$ for $R_V=3.1$. 
  The range derived above may underestimate the error.  
  We do not use the extinction for $R_C$ and $I_C$ as the constraint: 
  if we would take their 1.5 $\sigma$ error range for FM07, all 
  constraints discussed in this work will be unchanged.}
Figure 1(a) shows that the CCM curve with $R_V = 2.75-3.60$ and the 
dust-model calculation of WD01 (which gives $R_V = 3.1$) are both 
consistent with the data of FM07 in the optical and the UV regions.   
In the NIR region, however, the expanded figure, Figure 1(b), indicates 
that both curves are off from the FM07 data by $\approx$1 $\sigma$ or 
more.  
This arises from the fact that CCM adopted the NIR extinction that 
follows the power-law $A_\lambda/A_V \propto \lambda^{-\gamma}$ with 
the index $\gamma = 1.61$ (Rieke \& Lebofsky 1985), whereas the FM07 
data are consistent with a steeper index, $\gamma = 1.84$.  
The steeper index (Martin \& Whittet 1990) has gained more supports in 
the recent work: for example, Fitzpatrick \& Massa (2009) reported 
the index 1.78--2.0 and that it varies according to the line of sight 
[see also Fritz et al.\ (2011) for a summary of the recent work]. 
The slope changes between $I_C$ and $J$.

%%%%%%%%%%%%%%%%%%%%%%%%%%%%%%%%%%
\begin{figure}
%\epsscale{0.6}
\epsscale{1.1}
\plotone{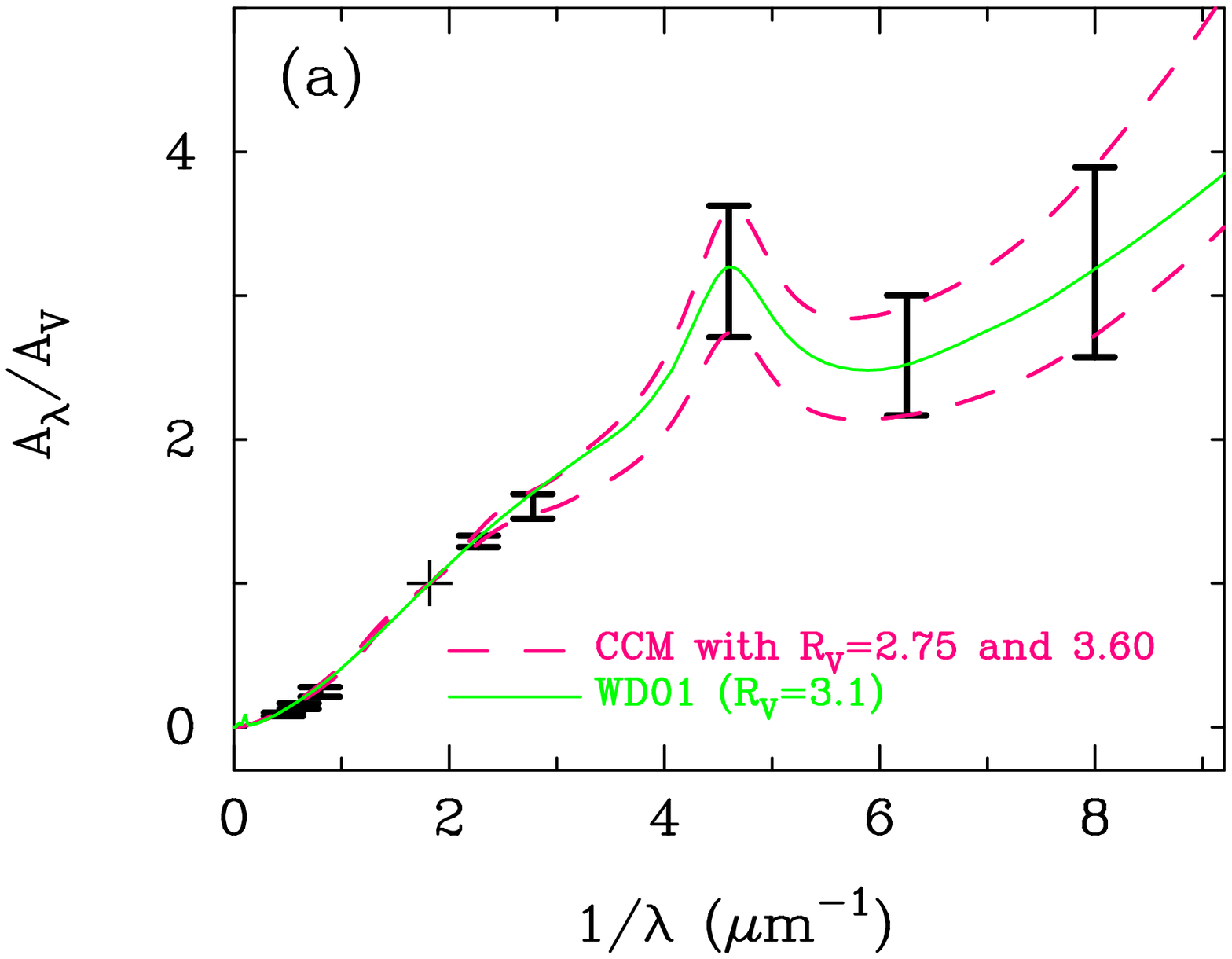}
\vspace{0.8 cm}
\plotone{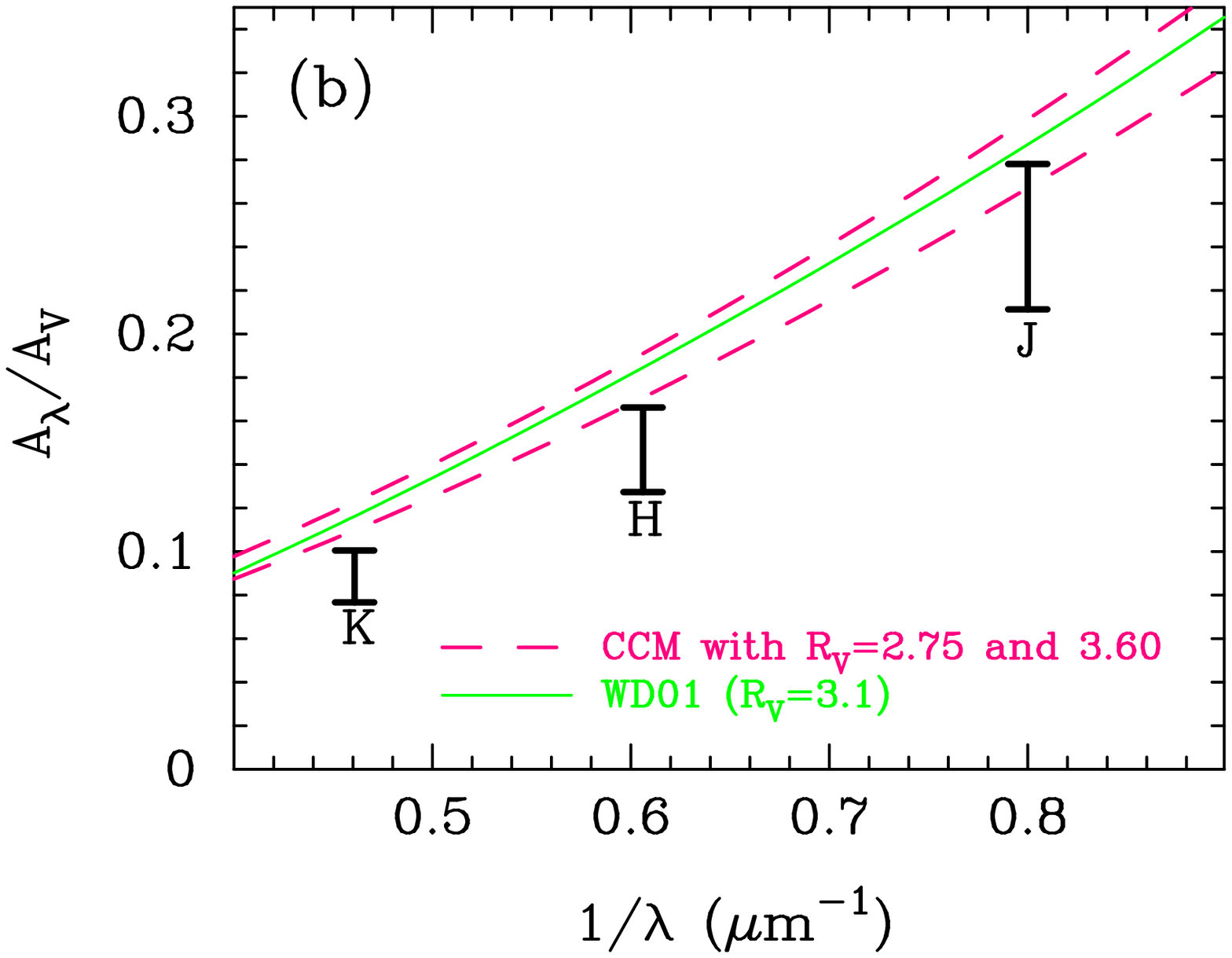}
\caption{ 
(a) Interstellar extinction curves from the 328 stars in the Milky 
Way from FM07. 
Our 1 $\sigma$ (thick solid bars) estimates are denoted at our 
reference wavelengths. 
The plus symbol shows the $V$ band used as the normalisation.  
(b) The expanded figure for the extinction in NIR. 
The bars are 1 $\sigma$.  
The extinction curves of the CCM formula with $R_V =$ 2.75 [the upper 
dashed curve in (a) and the lower dashed curve in (b)] and $R_V =$ 
3.60 [the lower dashed curve in (a) and the upper dashed in (b)], and 
the model of WD01 ($R_V =$ 3.1, the solid curves) are shown.
\label{fig1}}
\end{figure}
%%%%%%%%%%%%%%%%%%%%%%%%%%%%%%%%%%

We take the FM07 data as our prime choice, while tentatively retaining
the possibility that the NIR power index is moderate, as with CCM and
WD01.  
We also consider this possibility and study the implication on the 
dust model.  
We take the range of the CCM curves with $R_V =$ 2.75--3.60, which 
correspond to 1 $\sigma$ of the extinction data of FM07 in the UV and 
optical range, as seen in Figure 1(a).  
Our 1 sigma range adopted for NIR covers the variation of the index 
reported by Fitzpatrick \& Massa (2009), while CCM and WD01 are beyond 
1 $\sigma$.
The difference in the NIR slope leads to a significant difference in the
conclusion concerning the model.

In addition, we consider the absolute amount of extinction, or reddening.  
Bohlin, Savage, \& Drake (1978) obtained for 75 stars 
$N_{\rm H}/E(B-V) = 5.8 \times 10^{21}$ cm$^{-2}$ mag$^{-1}$, which is 
widely adopted in the literature, and claimed that the data for 
different lines of sight rarely fall beyond the lines 1.5 or 1/1.5
times the value indicated in this expression.  
Actually, we see in their figure (their Figure 2b) that about 85 \% of 
stars are located between the two lines. 
(Most of the deviants are towards the smaller $N_{\rm H}$ side.)  
The $N_{\rm H}$ value includes a 25 \% contribution from H$_2$ molecules.  
Unfortunately, they have not given the error or the dispersion of the 
fit.  
So we have re-fitted their data to obtain the 1 $\sigma$ error.  
We have also tried to include the recent enlarged data set compiled by  
Gudennavar et al.\ (2012).  
After the selection similar to that in Bohlin et al.\ (1978) we obtain, 
using 174 data,
\begin{eqnarray}
N_{\rm H}/E(B-V) = (5.7 \pm 1.7) \times 10^{21} 
~~{\rm cm}^{-2}~{\rm mag}^{-1},
\end{eqnarray}
where the error stands for the dispersion of the fit.  
This is in good agreement with the original Bohlin et al.\ (1978) 
result, including the size of the dispersion we re-estimated from their
data ($\pm 1.7$).

%%%%%%%%%%%%%%%%%%%%%%%%%%%%%%%%%%
\begin{deluxetable}{llccc}
\tablewidth{0pt}
\tablecaption{Extinction Data in the Small
Magellanic Cloud. }
\tablehead{
\multicolumn{2}{l}{Reference Wavelength} &
\multicolumn{1}{c}{$1/\lambda$} &
\multicolumn{2}{c}{($A_\lambda/A_V$)$_{\rm obs}$}  \\
\multicolumn{1}{c}{($\mu$m)} &  &
\multicolumn{1}{c}{($\mu$m$^{-1}$)} &
\multicolumn{1}{c}{upper} &
\multicolumn{1}{c}{lower} 
}
\startdata
  & & & & \\
0.125  & (far-UV rise) & 8.00 & 6.752 & 4.839 \\
0.16   & (far-UV dip)  & 6.25 & 5.005 & 3.813 \\
0.2175 & (UV Bump)     & 4.60 & 3.946 & 2.772 \\
0.36   & ({\it U} band)   & 2.78 & 1.892 & 1.521 \\
0.44   & ({\it B} band)   & 2.27 & 1.488 & 1.303 \\
0.55   & ({\it V} band) &  1.82 & 1 & 1 \\
1.25   & ({\it J} band)   & 0.80 & 0.324 & 0.108 \\
1.65   & ({\it H} band)   & 0.61 & 0.184 & 0.000 \\
2.17   & ({\it K} band)   & 0.46 & 0.060 & 0.000 \\
\enddata
\end{deluxetable}
%%%%%%%%%%%%%%%%%%%%%%%%%%%%%%%%%%

%%%%%%%%%%%%%%%%%%%%%%%%%%%%%%%%%%
\begin{figure}
%\epsscale{0.6}
\epsscale{1.1}
\plotone{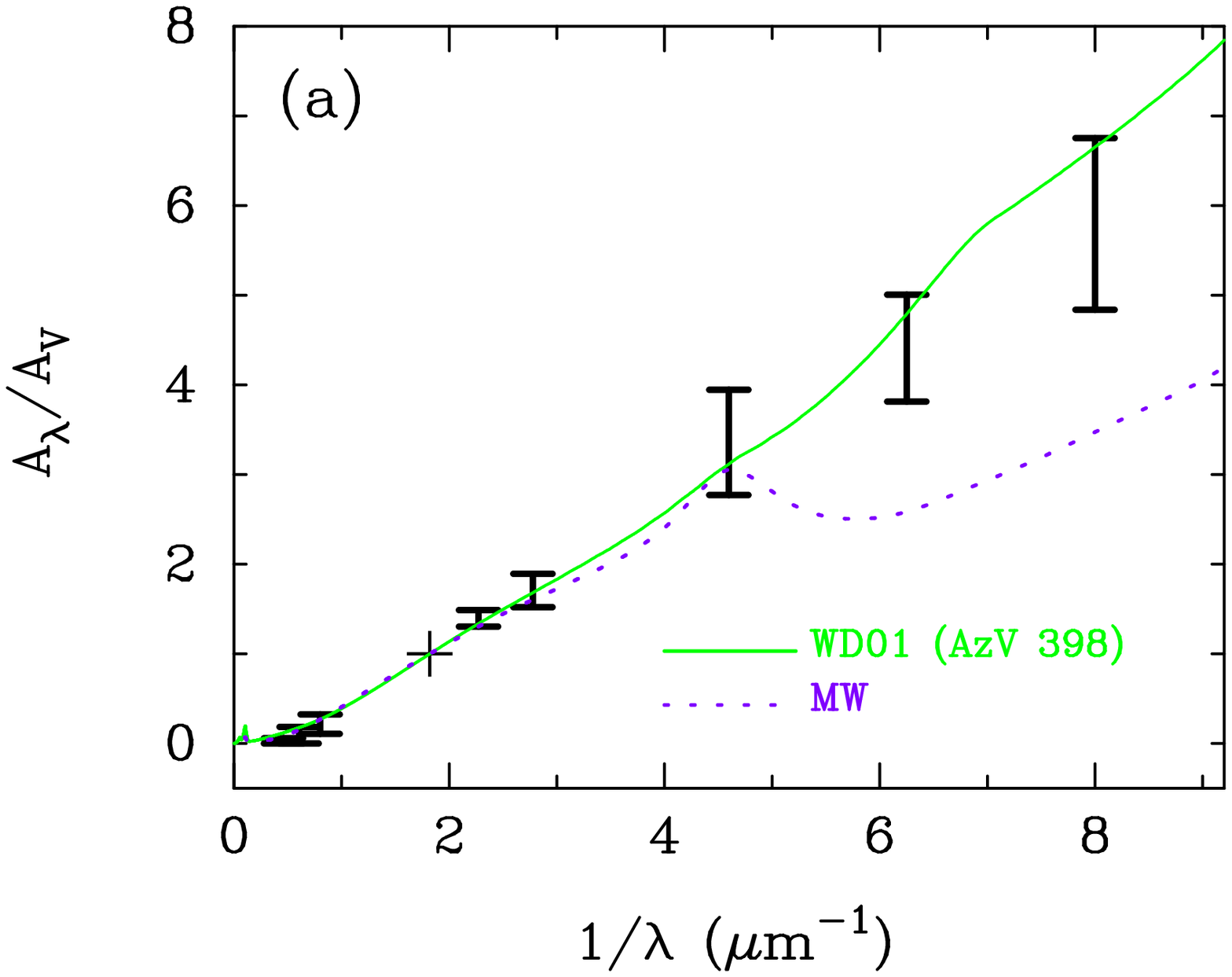}
\vspace{0.8 cm}
\plotone{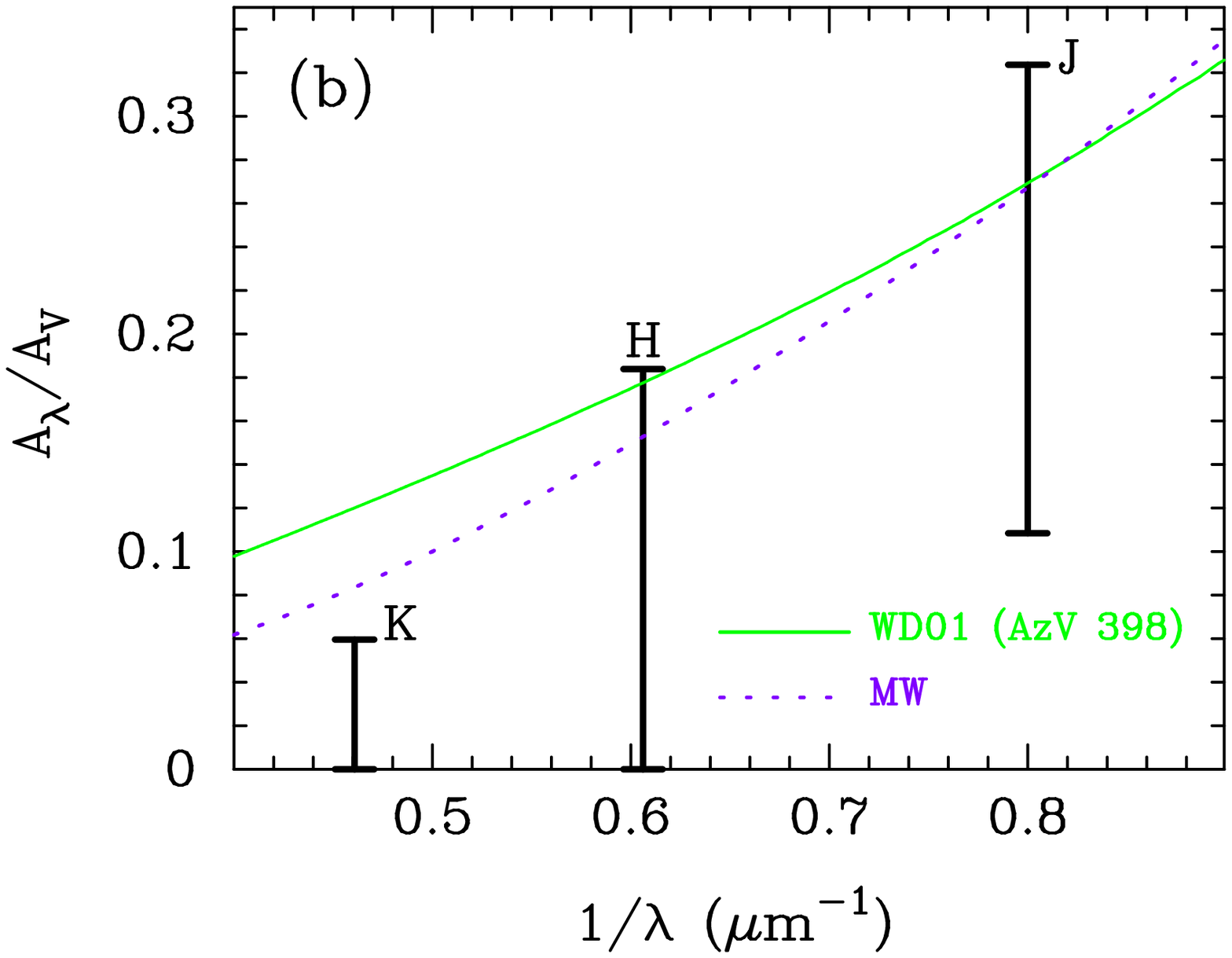}
\caption{ 
(a) The extinction curves for the SMC derived from five stars by
G03, with 1 $\sigma$ (thick solid bars) significance denoted at the
reference wavelengths.  
The plus symbol corresponds to the $V$ band.
(b) The expanded figure for the NIR extinction.  
The extinction curve given by WD01 (for AzV 398) is shown by the solid 
line. 
The extinction curve for the MW, taken from Figure 3 below (our model), 
is also added by thin dotted curves for comparison.
\label{fig2}}
\end{figure}
%%%%%%%%%%%%%%%%%%%%%%%%%%%%%%%%%%

%%%%%%%%%%%%%%%%%%%%%%%%%%%%%%%%%%%%%%%%%%%%%%%%%%%%%%%%%%%%%%%%%%%%%%
\subsection{Extinction in the Small Magellanic Cloud}
%%%%%%%%%%%%%%%%%%%%%%%%%%%%%%%%%%%%%%%%%%%%%%%%%%%%%%%%%%%%%%%%%%%%%%

Only a handful of sightlines are studied for the extinction curve for  
the SMC.
Gordon et al.\ (2003, hereafter G03) presented the curve towards 5 
stars, AzV 18, AzV 23, AzV 214, AzV 398, and AzV 456 using the
traditional reddened-comparison pair of stars.
More recent result of Cartledge et al.\ (2005) is consistent with G03.

As known, the bump at 2175 \AA~is not apparent except for the
sightline towards AzV 456, which is located in a \lq wing' region of
quiescent star formation, unlike the other four that pass through the
star-forming bar of the SMC.  
We include the extinction curve towards AzV 456 in our consideration.

For our measure we define, because of the paucity of data, the \lq 1
$\sigma$ ranges' of the extinction to be the maximum and the minimum
of the five curves at each reference wavelength.  
If the lower value becomes negative, it is set to zero.  
The ranges of extinction are given in Table 2 and plotted in Figure 2.

The recent analysis for the column density of neutral hydrogen 
$N_{\rm HI}$ is by Welty et al.\ (2012).  
The abundance of H$_2$ molecules is estimated to be 2 \% of hydrogen, 
which is compared to 25 \% for the MW (Bohlin et al.\ 1978), giving 
an example of molecular formation depending on environment.  
Their estimate is consistent with that of Tumlinson et al.\ 
(2002)\footnote{The estimate of Cartledge et al.\ (2005) for H$_2$ for 
  some of the 5 stars wildly varies from 5 \% to 50 \%.  We quote that 
  the cosmic global abundance of $2 N({\rm H}_2)/N({\rm H})$ is 
  inferred to be 0.30 (Fukugita 2011).}, 
$2 N_{\rm H_2} / N_{\rm H} = 1 {+0.5 \atop -0.2}$ \%.  
Welty et al.\ (2012) have given
\begin{eqnarray}
N_{\rm H}/E(B-V) =
2.3{+2.8 \atop -1.3}\times 10^{22} ~~{\rm cm}^{-2}~{\rm mag}^{-1}.
\end{eqnarray}
This is larger than the MW value given in Equation (1) by a factor of
$4.0 \pm 2.4$.  
Russell \& Dopita (1992) give a summary of the elemental abundance 
for the SMC, which indicates that its metallicity is 5.6 times smaller 
than the solar.  
We note that the abundance in SMC is poorly known.
The inferred dust abundance per hydrogen of the SMC, relative to the 
MW, is consistent with, or somewhat less suppressed compared with, the 
inferred heavy element abundance.

%%%%%%%%%%%%%%%%%%%%%%%%%%%%%%%%%%%%%%%%%%%%%%%%%%%%%%%%%%%%%%%%%%%%%%%
\section{Dust model}
%%%%%%%%%%%%%%%%%%%%%%%%%%%%%%%%%%%%%%%%%%%%%%%%%%%%%%%%%%%%%%%%%%%%%%%

With spherical particles uniformly distributed in interstellar space
the total extinction, $A_\lambda$, along the line of sight $l$, is
\begin{eqnarray}
A_\lambda = 1.086 \sum_j \int dl 
\int^{a_{{\rm max},j}}_{a_{{\rm min},j}} 
\pi a^2 Q^{\rm ext}_{\lambda,j}(a) n_j(a) da,
\end{eqnarray}
where $Q^{\rm ext}_{\lambda,j}(a)$ is the extinction efficiency
defined as the ratio of the extinction cross section
$\sigma_{\lambda,j}(a)$ at wavelengths $\lambda$ to the geometric
cross section $\pi a^2$ for grain species $j$ and is calculated using
Mie scattering with the laboratory optical data of dielectric
constants: 
$n_j(a)da$ is the number density of the grain of species $j$ with 
radius between $a$ and $a+da$.

The ``standard composition'' of dust is graphite and silicate.
Following DL84 we take ``astronomical silicate'' (we simply call it
silicate unless otherwise stated) with the composition MgFeSiO$_4$.  
We assume the simple approximation that dust grains are
bare refractory particles without substructure.  
For graphite, we calculate $Q_{\lambda, {\rm gra}}^{\rm ext}$ using 
the usually adopted $\frac{1}{3}-\frac{2}{3}$ approximation for the 
dielectric constant components perpendicular and parallel to the basal 
plane (e.g., DL84) to represent anisotropy.

In order to see their possible importance we also consider other
refractory components and carbonaceous materials, which are likely to
condense into grains, in so far as their optical properties are known.  
We consider 10 grain species, graphite (gra; optical data from Draine
2003b), glassy carbon (glC; Edoh 1983), amorphous carbon (amC; Zubko
et al.\ 1996), silicon carbide (SiC; Choyke \& Palik 1985),
astronomical silicate (asil; Draine 2003b) of the chemical composition 
of olivine (Mg$_x$Fe$_{1-x}$SiO$_4$) with $x \approx 1$, forsterite
(Mg$_2$SiO$_4$; Semenov et al.\ 2003), pure iron (Fe, Semenov et
al.\ 2003), magnetite (Fe$_3$O$_4$; Triaud)\footnote{A.\ Triaud,
  http://www.astro.uni-jena.de/Laboratory/OCDB/
  oxsul.html}, 
troilite (FeS; Semenov et al.\ 2003), and corundum (Al$_2$O$_3$; 
Toon et al.\ 1976).  
We do not take enstatite (MgSiO$_3$).  
One may expect that the optical property of MgSiO$_3$ is not much 
different from that of olivine, and its contribution may not disturb 
much the calculation with astronomical silicate.\footnote{The 
  imaginary part of the presently available refractive index of 
  enstatite (Dorschner et al.\ 1995) deviates largely from that of 
  forsterite in the UV region, while the real part differs little.  
  The Q factor of enstatite differs little from that of forsterite. 
  We do not treat enstatite separately in this paper.}  
We ignore SiO$_2$, since its $Q$ factor is small. 
Inclusion of SiO$_2$ only increases the amount of Si locked in dust 
grains.  
The Ca- and Ti-bearing grains may constitute some components of dust, 
but they are not considered because of their small cosmic abundance.

We do not consider polycyclic aromatic hydrocarbons (PAHs) separately.
The bump at 2175 \AA~in the Galactic extinction curve can be accounted
for either with or without PAH in so far as small graphite grains
(molecules with the effective radius of $\lesssim$200 \AA) are
included (Stecher \& Donn 1965; Joblin et al.\ 1992; Clayton et
al.\ 2003b).

Dust grains dominated by a single size do not give the observationally
obtained extinction curve.  
Some distribution over the size is necessary.  
A typical distribution that is known to work is the power law, which 
we also take here:
\begin{eqnarray}
n_j(a) = n_{\rm H} K_j (a/a_0)^{-q_j},
\end{eqnarray}
where $n_{\rm H}$ is the hydrogen number density, $K_j$ is the
fraction of species $j$, and $a_0$ is a size of the normalisation.  
We limit the range to $a_{{\rm min},j} \le a \le a_{{\rm max},j}$.  
We assume the same grain-size distribution independent of grain 
species.

We take $q_j$, $a_{{\rm max},j}$, and the condensed fraction
$f_{i,j}$, i.e., the fraction of element $i$ contained in grain
species $j$, as parameters, and find a set of the parameters that
satisfy 1-$\sigma$ of extinction data at {\it all}~ 9 reference
wavelengths.  We also study the dependence on the parameter 
$a_{{\rm min},j}$, but the fit varies little in so far as this minimum size
is smaller than 0.005 $\mu$m.  The minimum size $a_{\rm min}$
affects the extinction curve in the way that increasing it diminishes
the rise in the far UV for wavelengths roughly shorter than $O(2\pi)
\times a_{\rm min}$.  
With our choice of $a_{\rm min}=50$ \AA~it affects little the curve 
for $\lambda >$ 1150--1250 \AA~we consider.
The increasing variation of $a_{\rm min}$ also affects the 2175
\AA~bump caused by graphite, reducing the hump, but the resulting
change of the extinction curve from the decreasing variation of
$a_{\rm min}$ from 0.005 $\mu$m to zero is small, typically $<$1/3 the
1 $\sigma$ error range.  
The parameter $a_{\rm min}$ is not well determined if it is left as a 
free parameter, unless the extinction data of shorter wavelengths are 
used.  
So we fix $a_{\rm min}$ at 0.005 $\mu$m\footnote{Considering a possible 
  importance of small grains beyond the power law in the IR emission, 
  we examined the possibility that some non-negligible amounts of grains 
  are distributed below $a_{\rm min}$. 
  The presence of such grains, say at 0.001 or 0.005 $\mu$m, modifies 
  most characteristically the UV slope, making it steeper at wavelengths 
  shorter than 2000\AA~and the hump of 2175\AA~larger. 
  In so far as these extra components are less than 15\% in mass, our 
  resulting extinction curves are not changed beyond our reference 
  errors.}.  
The change of lower cutoff affects little the determination of other 
parameters, unless $q \gtrsim 4$, which is deep in the region not allowed 
in our study.  
It increases the total amount of mass, say, by 10 \% if $a_{\rm min}$ 
is reduced from 0.005 $\mu$m to 0.001 $\mu$m.  
A possible increase of the lower cutoff for graphite is considered for 
SMC dust separately.  
The extinction curve we refer to is $A_\lambda/A_V$ normalised in the 
$V$ band:
\begin{eqnarray}
\frac{A_\lambda}{A_V} = \frac{\sum_j K_j A_{\lambda,j}}
{\sum_j K_j A_{V,j}},
\end{eqnarray}
where $A_{\lambda,j}$ is the component contribution to Equation (3).

We also calculate
\begin{eqnarray}
\frac{E(B-V)}{N_{\rm H}} = 
1.086\sum_j K_j  \int^{a_{{\rm max},j}}_{a_{{\rm  min},j}} da 
(\sigma_{B,j} - \sigma_{V,j}) \left( \frac{a}{a_0} \right)^{-q_j}.
\end{eqnarray}
The amount of reddening per hydrogen leads to the constraint on the
abundance of dust that can be compared with other 
elemental abundance estimates.  
Assuming that the abundance does not vary from place to place, we take 
as standard for the MW the solar elemental abundance estimated by 
Grevesse \& Sauval (1998) (hereafter GS98).\footnote{We remark that 
  GS98 agrees with the earlier table of
  Anders \& Grevesse (1989) up to oxygen, for which GS98 value is
  lower by 0.1 dex.}  
Asplund et al.\ (2009), using the 3D calculation, claimed the solar 
abundance generally lower than that of GS98 by as much as 30 \%, 
especially for C, O and a few others.  
A comparable 3D calculation (Caffau et al.\ 2011), however, has given
an abundance that is lower than GS98 only by 10 \%.

The elemental abundance is generally not tightly converged among the
authors. 
For instance, for the carbon abundance, which is of one of our major 
concerns, the GS98 value C/H $= (3.3 \pm 0.5)\times 10^{-4}$ may be 
compared with 
$3.62 \times 10^{-4}$ of Anders \& Grevesse (1989),
$2.69\times 10^{-4}$ of Asplund et al.\ (2009),
 $3.16 \times 10^{-4}$ of Caffau et al. (2011),
(1.9--2.9)$\times 10^{-4}$ of Cardelli et al.\ (1996), 
$2.14 \times 10^{-4}$ of Nieva \& Przybilla (2012), 
$2.45 \times 10^{-4}$ of Lodders (2010), and so forth. 
The iron abundance, for which GS98 gives 
$\log ({\rm Fe/H}) + 12=7.50\pm 0.05$, too, varies between 7.45 and 7.66. 
Keeping these uncertainties in mind, we take as our default GS98, which 
leads to a satisfactory solar structure.

For the SMC we take the composition given by Russell \& Dopita (1992),
with which total metallicity is 1/5.6 times the solar.  
The abundance of refractory elements, Mg, Si, and Fe, is smaller than 
their MW values by factors, 1/3.3 to 1/4.6.  
The abundance of O and C in the SMC seems more strongly depressed, by 
1/(6.2--6.3) times, compared with the solar abundance.  
Taken literally, this leads to the significant result that the ratio of 
carbonaceous material to silicate in the SMC is 50 \% lower than the 
corresponding value for the MW. 
We must remember, however, that the elemental abundance for SMC is 
probably more uncertain than for MW.

%%%%%%%%%%%%%%%%%%%%%%%%%%%%%%%%%%%%%%%%%%%%%%%%%%%%%%%%%%%%%%%%%%%%%%%
\section{Results for MW dust}
%%%%%%%%%%%%%%%%%%%%%%%%%%%%%%%%%%%%%%%%%%%%%%%%%%%%%%%%%%%%%%%%%%%%%%%

%%%%%%%%%%%%%%%%%%%%%%%%%%%%%%%%%%%%%%%%%%%%%%%%%%%%%%%%%%%%%%%%%%%%%%%
\subsection{Single grain species}
%%%%%%%%%%%%%%%%%%%%%%%%%%%%%%%%%%%%%%%%%%%%%%%%%%%%%%%%%%%%%%%%%%%%%%%

We first study the extinction curve with a single grain species.  
We consider the allowed region in $q$ and $a_{\rm max}$ plane, with 
which the dust model satisfies the 1 $\sigma$ ranges of the FM07 
extinction curve.  
We find that there is no overlap among the three regions derived from 
the UV group (0.125 $\mu$m, 0.16 $\mu$m, and 0.2175 $\mu$m), the UBV 
group, and the NIR group ({\it J}, {\it H}, and {\it K}) for any 
species of grains we considered.  
The regions required for UV and UBV groups are always disjoint. 
It often happens that no consistent parameters exist for {\it J}, 
{\it H}, and {\it K}, the NIR passbands alone.  
We cannot make a dust model which explains the extinction curve over 
the wide range of wavelength with only a single grain species.  
For instance, astronomical silicate gives too steep a rise in the far UV 
if $a_{\rm max}$ is chosen to account for the optical extinction curve,
and the predicted NIR extinction is too small by a factor of 3 to 4. 
We expect that these problems are offset by introducing carbonaceous 
grains, which give a milder rise in the far UV and have a larger 
scattering efficiency in the NIR bands, and we suppose that the mixture 
of these two species would give the correct extinction curve.

%%%%%%%%%%%%%%%%%%%%%%%%%%%%%%%%%%
\begin{figure}
%\epsscale{0.6}
\epsscale{1.1}
\plotone{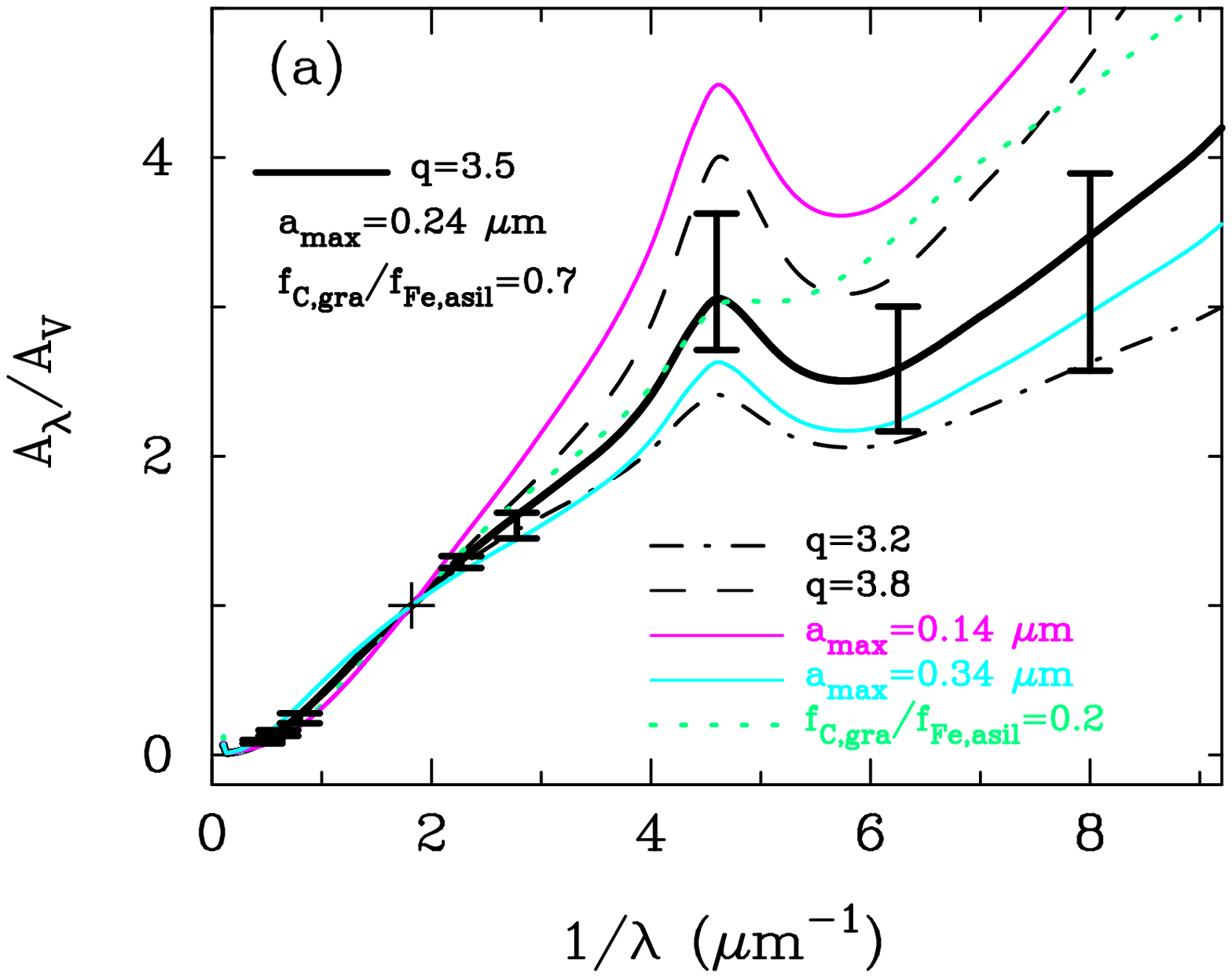}
\vspace{0.8 cm}
\plotone{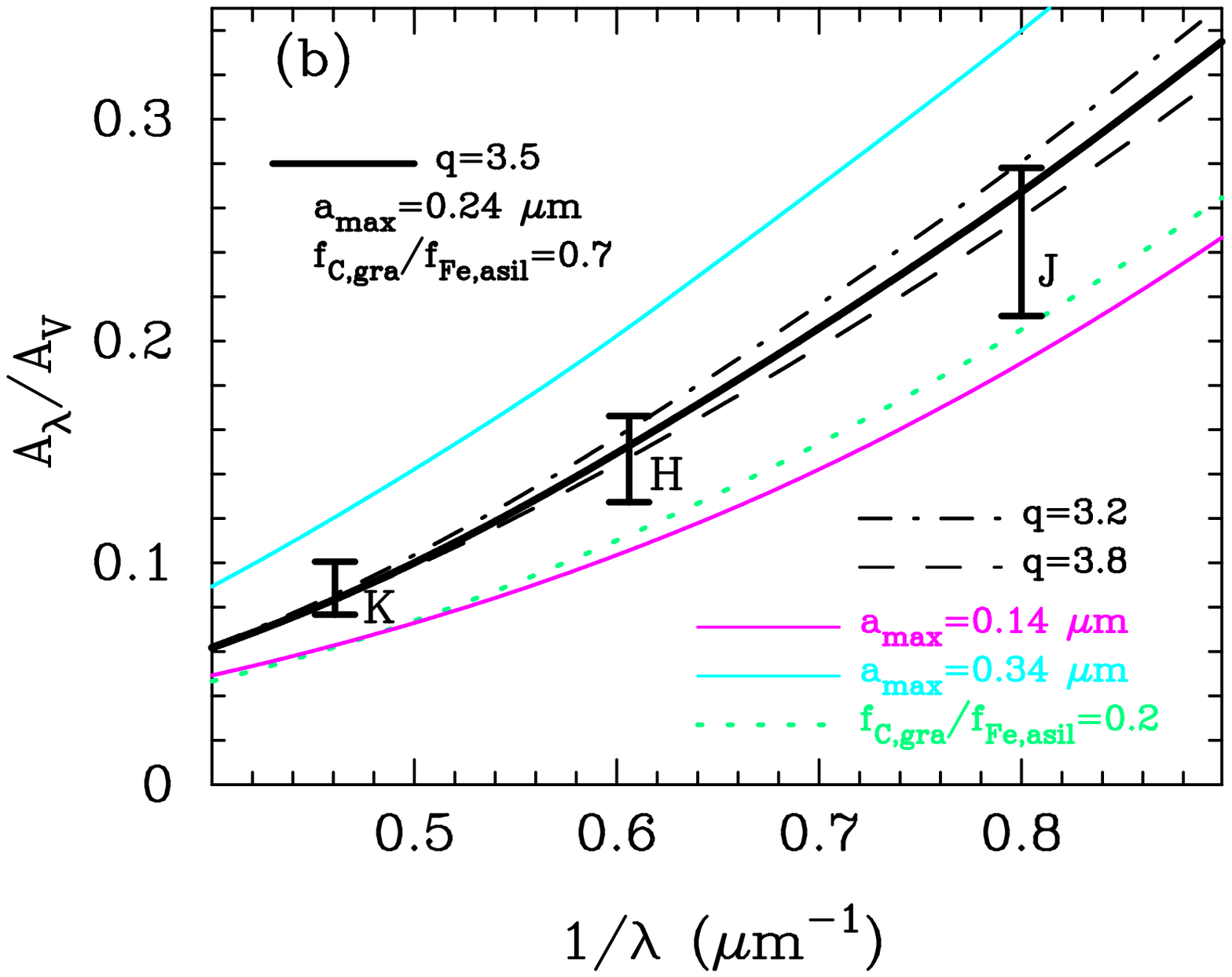}
\caption{
(a) Typical extinction curves from the graphite-silicate models, 
and (b) the expanded figure of (a) for NIR wavelengths.
The error bars stand for the observed 1 $\sigma$ ranges.
The thick solid curve is with $q=3.5$, $a_{\rm max}= 0.24$ $\mu$m,
and $f_{\rm C,gra} / f_{\rm Fe,asil} =0.7$,
taken as our fiducial. 
The other curves denoted by thin curves are examples where one of
the three parameters is shifted to demonstrate the response to the 
parameter.
The legend shows the parameters when changed from the fiducial choice.
\label{fig3}}
\end{figure}
%%%%%%%%%%%%%%%%%%%%%%%%%%%%%%%%%%

%%%%%%%%%%%%%%%%%%%%%%%%%%%%%%%%%%
\begin{deluxetable}{lllll}
\tablewidth{0pt}
\tablecaption{Extinction Data $A_\lambda/A_V$ for the Milky Way.
Our model extinction is compared with 
those of CCM and WD01}
\tablehead{
\multicolumn{2}{l}{Wavelength} &
\multicolumn{3}{c}{$A_\lambda/A_V$} \\
\multicolumn{1}{c}{($\mu$m)} &  &
\multicolumn{1}{c}{our work} &
\multicolumn{1}{c}{CCM} &
\multicolumn{1}{c}{WD01}  \\
\multicolumn{1}{c}{} &  &
\multicolumn{1}{c}{} &
\multicolumn{1}{c}{($R_V = 3.1$)} &
\multicolumn{1}{c}{($R_V = 3.1$)} 
}
\startdata
  & & & & \\
0.3531 & ($u$) & 1.633 & 1.584 & 1.660 \\
0.365 & ($U$) & 1.583 & 1.557 & 1.608 \\
0.44 & ($B$) & 1.306 & 1.325 & 1.318 \\
0.4627 & ($g$) & 1.232 & 1.243 & 1.246 \\
0.55 & ($V$) & 1.000 & 1.000 & 1.000 \\
0.6140 & ($r$) & 0.866 & 0.884 & 0.859 \\
0.66 & ($R_{\rm c}$) & 0.785 & 0.812 & 0.774 \\
0.7467 & ($i$) & 0.660 & 0.678 & 0.646 \\
0.81 & ($I_{\rm c}$) & 0.582 & 0.583 & 0.571 \\
0.8887 & ($z$) & 0.498 & 0.489 & 0.493 \\
1.25 & ($J$) & 0.267 & 0.282 & 0.287 \\
1.65 & ($H$) & 0.153 & 0.180 & 0.185 \\
2.17 & ($K$) & 0.0836 & 0.116 & 0.117 \\
3.35 & (WISE1)    & 0.0348 & 0.0577 & 0.0513 \\
3.55 & (IRAC1)    & 0.0313 & 0.0525 & 0.0463 \\
4.44 & (IRAC2)    & 0.0213 & 0.0367 & 0.0297 \\
4.60 & (WISE2)    & 0.0201 & 0.0346 & 0.0280 \\
%
%5.73 & (IRAC3)    & 0.0147 & 0.0243 & 0.0196 \\
%
%8.0 & (IRAC4)    & 0.0215 & 0.0142 & 0.0289 \\
%
\enddata
\end{deluxetable}
%%%%%%%%%%%%%%%%%%%%%%%%%%%%%%%%%%

%%%%%%%%%%%%%%%%%%%%%%%%%%%%%%%%%%%%%%%%%%%%%%%%%%%%%%%%%%%%%%%%%%%%%%%
\subsection{Graphite-silicate dust model}
%%%%%%%%%%%%%%%%%%%%%%%%%%%%%%%%%%%%%%%%%%%%%%%%%%%%%%%%%%%%%%%%%%%%%%%

The two-component dust model, consisting of graphite and astronomical
silicate, has been widely taken for interstellar dust since MRN.  
Mg, Si, and Fe are similar in the cosmic number abundance (within 
20 \%).
If we assume that Si is all condensed into astronomical silicate, 
$f_{\rm Si,asil} = 1$, then $f_{\rm Mg,asil} = 0.93$ and 
$f_{\rm Fe,asil} = 1$, if Fe:Mg=1:1 (see Draine 2003a) with the GS98 
abundance.
Here iron is slightly (10 \%) deficient and extra Si and Mg may 
condense into forsterite.
Oxygen locked in silicate is $f_{\rm O,asil} = 0.26$. 
Carbon is also depleted in interstellar matter. 
Following Sofia et al.\ (2011), it is approximately 60--70 \%, i.e.,
\begin{equation}
f_{\rm C,gra} = 0.6 - 0.7
\label{eq:c-depletion}
\end{equation}
if all condensed carbon is in graphite. 
This is higher than their earlier estimate, 
30--40 \%.\footnote{Sofia et al.\ (2004) used only weak absorption 
  lines of carbon, while their updated analysis (2011) includes 
  strong absorption lines.}
The depletion given by Cardelli et al.\ (1996) is also consistent
with eq.({\ref{eq:c-depletion}) if the GS98 abundance is adopted.

The model attains good fits from far UV to NIR within 1 $\sigma$ of 
the data with
\begin{equation}
q = 3.5 \pm 0.2 {\rm ~~and~~} a_{\rm max} = 0.24{+0.10 \atop -0.05}.
\label{eq:fit}
\end{equation}
The range of $f_{\rm C,gra} / f_{\rm Fe,asil}$ is 0.25--2.23, 
allowing for $f_{\rm  Fe,asil} < 1$.
Figure 3 presents examples of the resulting extinction curve from 
our model at the central values of eq.(\ref{eq:fit}) with 
$f_{\rm C,gra} / f_{\rm Fe,asil} = 0.7$.  
We give in Table 3 numerical values of $A_\lambda/A_V$ for selected 
wavelengths (and compare them with the models of WD01 and CCM).
We underline the discrepancy between our model and WD01 (and also
CCM) increasing from the $J$ band longwards. 
It becomes 30\% for the $K$ band.
We draw several curves in addition that are somewhat away from the 
best fit value (say, about 1.5 $\sigma$ of the observation) to indicate 
how the curve shifts with the variation of $q$, $a_{\rm max}$, 
and $f_{\rm C,gra} / f_{\rm Fe,asil}$.
The mass density ratio of graphite to astronomical silicate is given 
by $\rho_{\rm gra} / \rho_{\rm asil} = 
0.73 f_{\rm C,gra} / f_{\rm Fe,asil}$. 
We note that our curve is in close match with the WD01 curve 
($R_V=3.1$) from the $U$ to $I_C$ passbands.\footnote{Our curve lies 
  at one sigma edges if we take the $R_C$ and $I_C$ data constructed 
  from the FM07 curves. 
  The constraints on $q-A_{\rm max}$ plane are unchanged if we take 
  1.5 $\sigma$ of $R_C$ and $I_C$. 
  Note that we do not use the $R_C$ and $I_C$ extinction data of FM07, 
  which are not observationally constrained.}
A significant departure starts from the $J$ band longwards.  
The role of graphite is, in addition to producing 2175 \AA~bump, 
to increase the extinction in NIR, which is too small with silicate
alone when normalised in the optical region. 
Graphite makes the rise in the far UV more moderate.

In Figure 4, we show in $q - a_{\rm max}$ plane the region where the 
model gives the 1 $\sigma$ ranges of the extinction for the choices of 
$f_{\rm C,gra}/f_{\rm Fe,asil}=$ 0.7 and 0.2.
The overlapping region is seen for UV, UBV, and NIR for 
$f_{\rm C,gra}/f_{\rm Fe,asil}=$ 0.7, validating the
graphite-silicate model.  
Such allowed regions disappear for 
$f_{\rm C,gra}/f_{\rm Fe,asil} \le 0.25$.

Figure 5 shows the abundances of carbon and silicon in grains 
necessary to account for the extinction curve relative to hydrogen. 
The allowed region extends in a belt, running from bottom left to top 
right, representing $f_{\rm C,gra}/f_{\rm Fe,asil} =$const. 
We also indicate the region allowed from $E(B-V)/N_{\rm H}$, which is
located near the centre of the belt (lightly shaded).
This leads to the elemental abundance carried by dust grains,
\begin{equation}
\log {\rm (C/H)}+12 = 8.4 \pm 0.3 ~{\rm and}~ 
\log {\rm (Si/H)}+12 = 7.6 \pm 0.4.
\end{equation}
We also indicate the abundance of GS98 for Si (7.55; horizontal dashed 
line) and C (8.52; vertical horizontal line) with the neighbouring 
shade showing various abundance estimates.
  
The abundances resulted from the interstellar extinction are consistent 
with other estimates.
If we take the GS98 value for the total abundance of C and Si 
including the gas phase, the figure, showing 
$\log {\rm (C/H)} + 12 \geq 8.27$, means
\begin{equation}
f_{\rm C,gra} \geq 0.56,
\end{equation}
for $f_{\rm Si,grain}=1$ assumed.  
This carbon fraction in graphite is consistent with the depletion 
estimated for interstellar matter quoted in 
eq.({\ref{eq:c-depletion}).\footnote{The carbon abundance required from 
  the extinction is still consistent with the lower abundance of 
  Asplund et al.\ (2009), but then the depletion must be as high as 
  $>$80 \%.}

%%%%%%%%%%%%%%%%%%%%%%%%%%%%%%%%%%
\begin{figure}
%\epsscale{0.6}
\epsscale{1.1}
\plotone{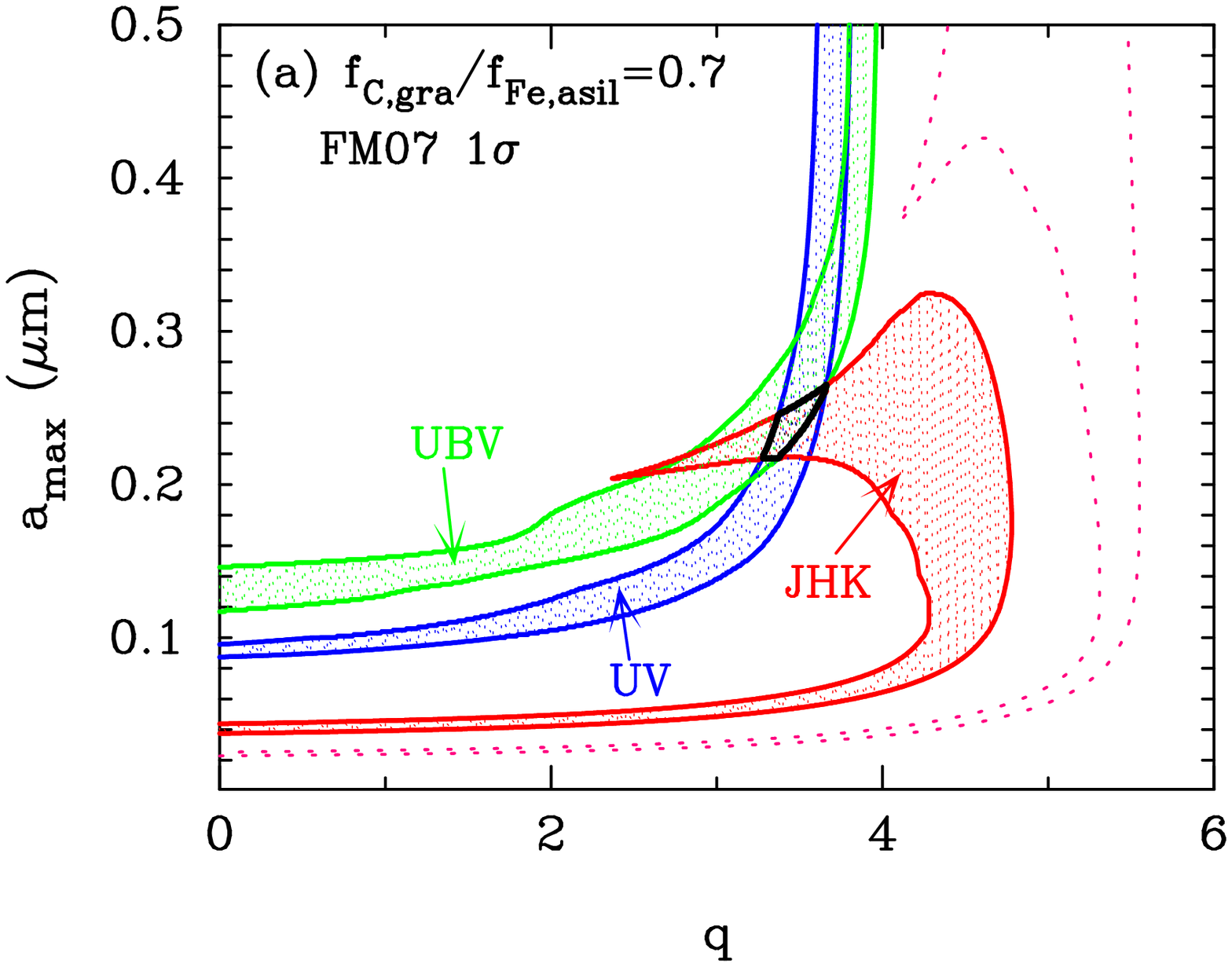}
\vspace{0.2 cm}
\plotone{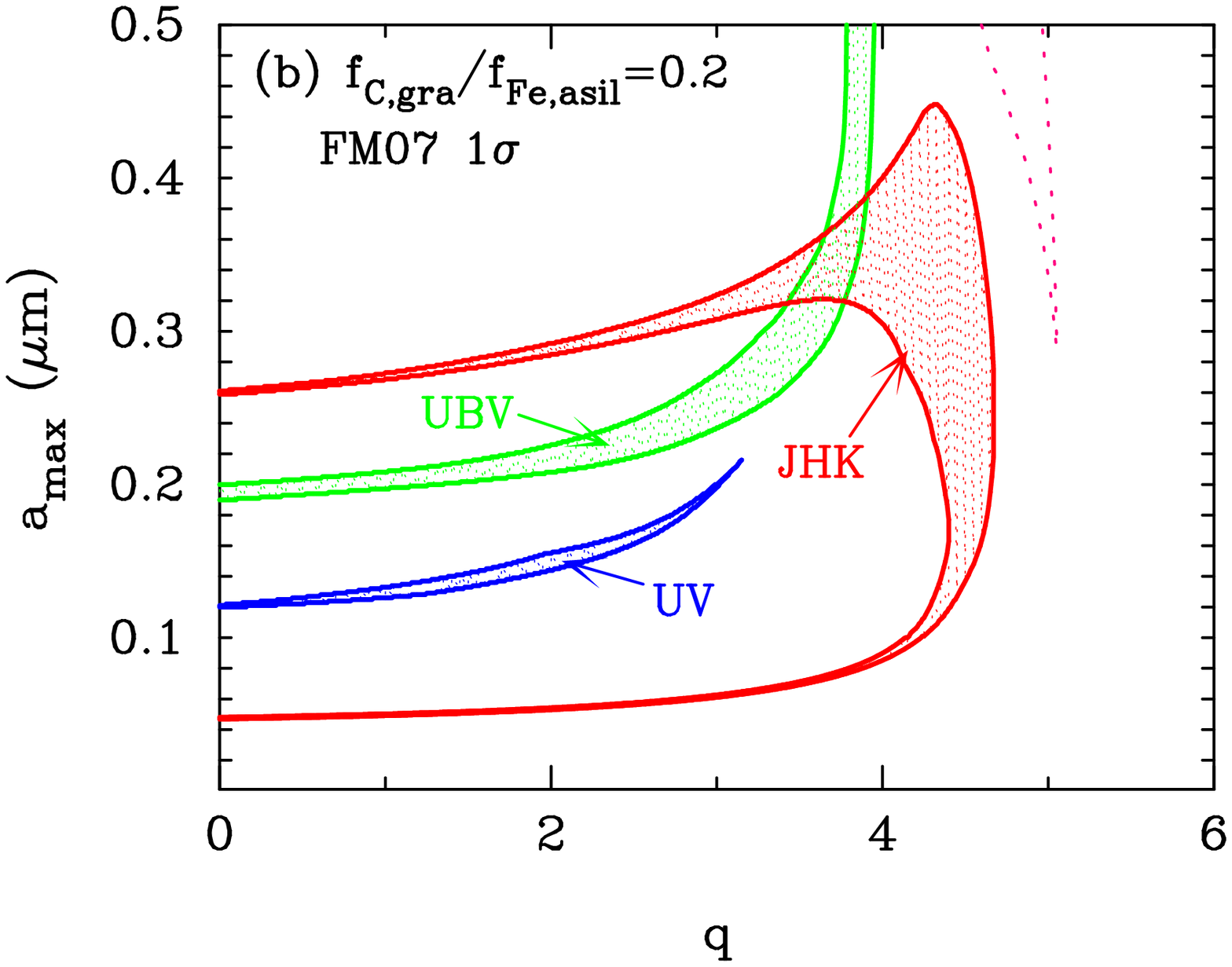}
\caption{
Allowed parameter regions for graphite-silicate models in $q$ and 
$a_{\rm max}$ plane for the 1 $\sigma$ range in the UV group (blue), 
the UBV group (green), and the NIR (JHK) group (red).  
We take $f_{\rm C,gra}/f_{\rm Fe,asil} = 0.7$ in (a) and 0.2 in (b),
the latter of which is a model out of 1 $\sigma$.  
Thin dotted curves show the parameters for the 1 $\sigma$ range in the 
NIR (JHK) group when the NIR wavelength dependent power is 
$\gamma=1.6$ as with CCM or WD01.
\label{fig4}}
\end{figure}
%%%%%%%%%%%%%%%%%%%%%%%%%%%%%%%%%%

A similar consideration is made for the 1 $\sigma$ ranges from the CCM 
formula, where the NIR extinction takes a smaller power index of the 
wavelength dependence.  
The range of $q$ and $a_{\rm max}$ required for the NIR extinction 
curve is shown in Figure 4 with dotted curves (those for UV and 
UBV remain unchanged).  
We see that the region allowed simultaneously for the three groups of
color bands does not exist.  
The region for NIR is always disjoint from those for UBV and UV.  
No graphite-silicate model is consistent with the observed extinction, 
if the NIR extinction power is $\approx$$\lambda^{-1.6}$, for grains 
with the power-law size distribution.

The steeper wavelength dependence for the NIR power thus looks more
easily accommodated from the model point of view.  
We remark that WD01 tweaked significantly the grain size distribution 
from the power law, separately for graphite and silicate, adding 
different components so that the graphite-silicate model becomes 
consistent with the CCM-like extinction.  
This is also true for the size distributions obtained by other authors 
(listed earlier) when the CCM-like extinction is reproduced with the 
graphite-silicate model.  
Let us comment that the solution of WD01 takes the Si, Mg, and Fe 
cosmic abundance larger by 30--50 \% than the GS98 cosmic value.

We note that the allowed regions in Figure 4 above stand for the ranges
consistent with the variation of dust properties along lines of sight
in the MW.  
They are well converged to narrow regions of $q$ and $a_{\rm max}$ 
indicated in eq.(\ref{eq:fit}), despite an apparently significant 
variation of extinction curves. 
The size of grains is similar; 
only a small variation of the size parameter could cause the difference 
in the extinction curve.

%%%%%%%%%%%%%%%%%%%%%%%%%%%%%%%%%%
\begin{figure}
%\epsscale{0.8}
\epsscale{1.12}
\plotone{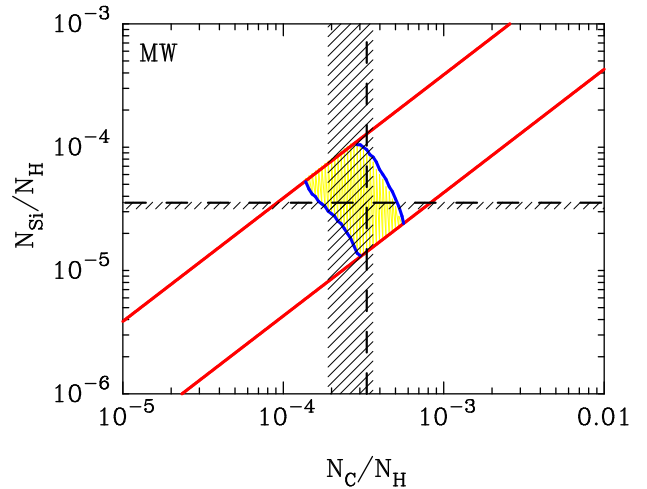}
\caption{
The abundance of C and Si in dust grains relative to hydrogen for 
the MW to satisfy 1 sigma range of the extinction curve. 
The range within the oblique belt is allowed from the extinction
curve, and further restricted range in the middle is from
$E(B-V)/N_{\rm H}$.  
The two horizontal and vertical dashed lines show the total abundance 
of Si and C from GS98, with the shaded regions corresponding to the 
range of various estimates.
\label{fig5}}
\end{figure}
%%%%%%%%%%%%%%%%%%%%%%%%%%%%%%%%%%

We summarise in Figure 6 the region allowed for $a_{\rm max}$ and $q$,
marginalising over the ratio of condensed fractions of graphite to
silicate for $f_{\rm C,gra}/f_{\rm Fe,asil} \ge 0.25$.  
The curves are inlaid for $f_{\rm C,gra}/f_{\rm Fe,asil} = $ 0.4, 0.7,
and 1.  
The 1 $\sigma$ allowed range is realised with $3.2 \le q \le 3.7$ and 
0.19 $\mu$m $\le a_{\rm max} \le$ 0.34 $\mu$m.  
With increasing $f_{\rm C,gra}$, $a_{\rm max}$ becomes smaller.  
The other set of curves is the constraint from $E(B-V)/N_{\rm H}$ for 
$f_{\rm C,gra} = $ 0.4, 0.7, and 1 with $f_{\rm Fe,asil} =1 $. 
Overlaps are seen between the two curves for 
$f_{\rm C,gra}/f_{\rm Fe,asil} = $ 0.7 and 1, but not for 0.4.

For our model that satisfies 1 $\sigma$ constraints of both extinction
curve and the $E(B-V)$ size we have the mass density of dust relative
to that of the hydrogen gas,
$\rho_{\rm dust} / \rho_{\rm H} = 1 / (122{+48 \atop -75})$.
The allowed range would be tightly constrained if the abundance is
fixed to GS98 and $f_{\rm Fe,asil} = 1$ is assumed:
\begin{equation}
\rho_{\rm dust} / \rho_{\rm H} = 1 / (122{+9 \atop -16}).
\label{eq:rho_dust}
\end{equation}
The mass extinction constant 
$K_{{\rm ext},\lambda} = A_\lambda/\Sigma_{\rm dust}$
for the $V$ band is
\begin{equation}
K_{\rm ext,V}=(3.7 \pm 0.5)\times 10^{4} {~~\rm mag~ cm^2~ g^{-1}},
\label{eq:K_ext}
\end{equation}
which is compared to $2.8\times 10^{4}$ mag cm$^2$ g$^{-1}$ of WD01.
(If the abundance constraints are removed, eq.(\ref{eq:K_ext}) 
becomes $(3.6 \pm 1.0) \times 10^{4}$ mag cm$^2$ g$^{-1}$.)
The ratio of the mass density of graphite to silicate is
$\rho_{\rm gra} / \rho_{\rm asil} = 0.51 {+0.22 \atop -0.10}$,
or in terms of the number of C and Fe atoms in grains,
$N_{\rm C} / N_{\rm Fe} = 7.3 {+3.2 \atop -1.4}$,
corresponding to eq.(\ref{eq:rho_dust}).

%%%%%%%%%%%%%%%%%%%%%%%%%%%%%%%%%%
\begin{figure}
%\epsscale{0.8}
\epsscale{1.12}
\plotone{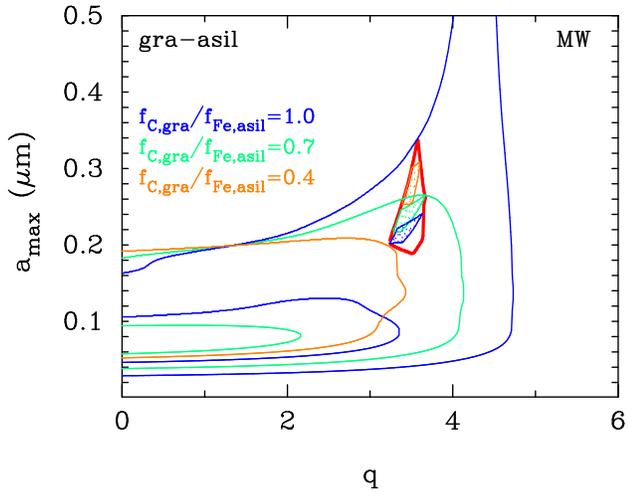}
\caption{ 
Allowed regions of $q$ and $a_{\rm max}$ with which the model 
satisfies the 1 $\sigma$ ranges of the observed extinction for the MW, 
marginalised over the graphite-to-silicate ratios 
$f_{\rm C,gra} / f_{\rm Fe,asil} \ge 0.25$.  
The three contours inlaid in the allowed region indicate those with 
$f_{\rm C,gra} / f_{\rm Fe,asil} = 1.0$, 0.7, and 0.4.  
The constraints from the reddening $E(B-V)/N_{\rm H}$ are shown for 
$f_{\rm C,gra} = 0.4$, 0.7, and 1.0 (with $f_{\rm Fe,asil} = 1.0$), 
drawn by thin (brown, green, and blue) contours.
\label{fig6}}
\end{figure}
%%%%%%%%%%%%%%%%%%%%%%%%%%%%%%%%%%

%%%%%%%%%%%%%%%%%%%%%%%%%%%%%%%%%%
\begin{figure}
%\epsscale{0.6}
\epsscale{1.1}
\plotone{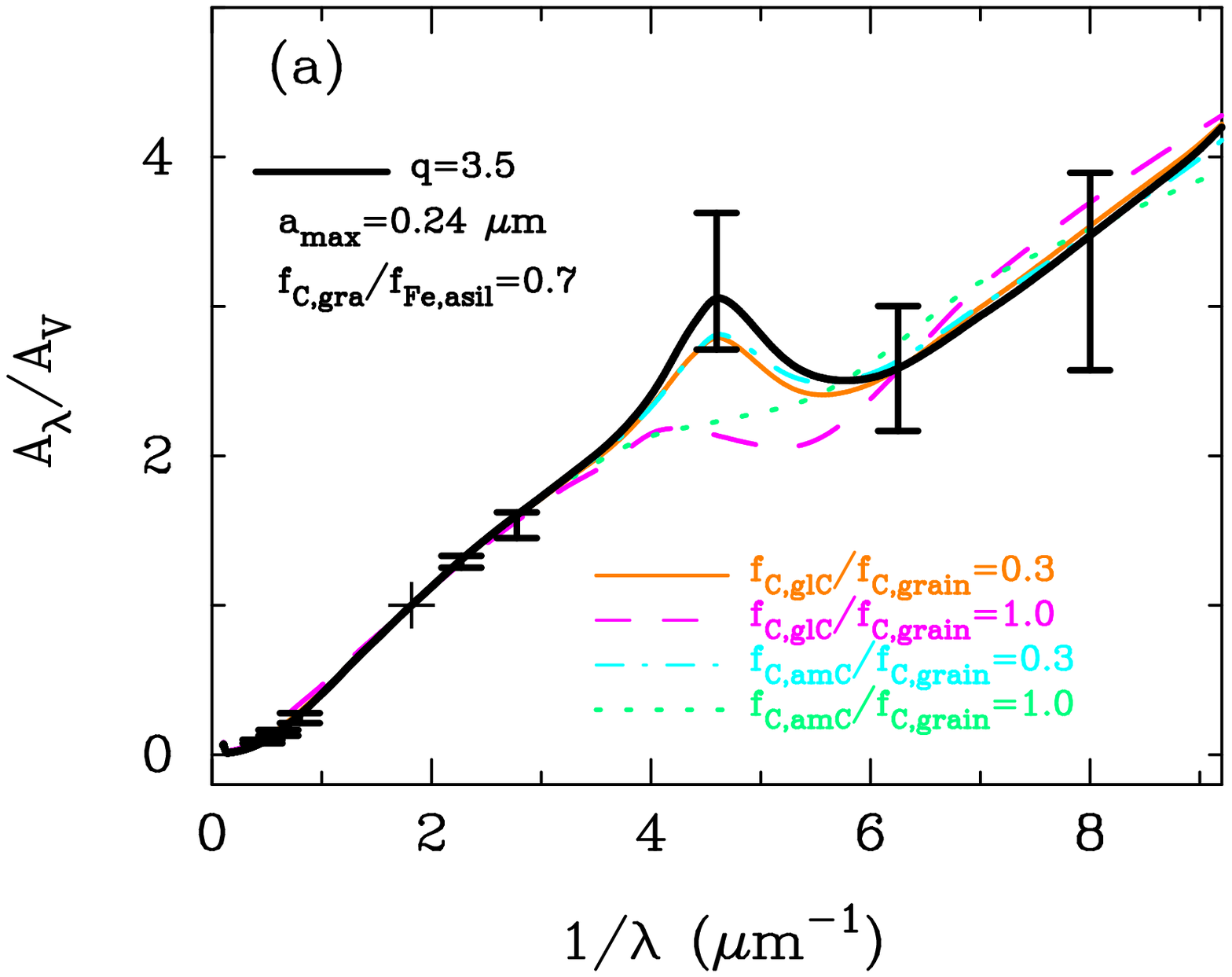}
\vspace{0.4 cm}
\plotone{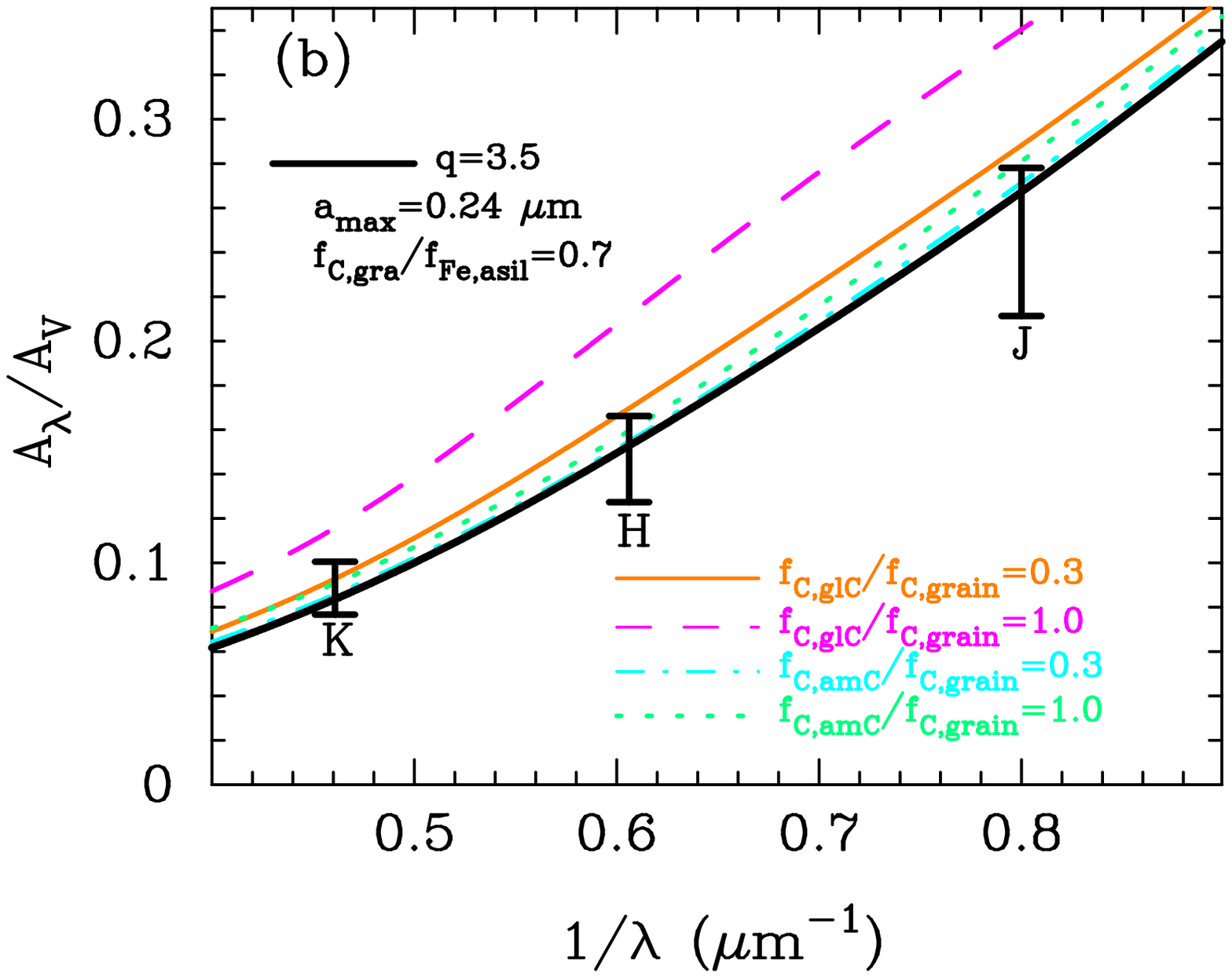}
\caption{ 
(a) Extinction curves from the graphite--glassy (or amorphous) 
carbon--silicate models with $q = 3.5$ and $a_{\rm max} = 0.24$ $\mu$m, 
and (b) the expanded figure of (a) for NIR, as compared with our 
fiducial graphite-silicate model (thick black curve).  
$f_{\rm C,grain} / f_{\rm Fe,asil} = 0.7$ is taken for all cases.  
The 1 $\sigma$ ranges from the observed extinction curves in
FM07 are indicated with error bars.
\label{fig7}}
\end{figure}
%%%%%%%%%%%%%%%%%%%%%%%%%%%%%%%%%%

%%%%%%%%%%%%%%%%%%%%%%%%%%%%%%%%%%%%%%%%%%%%%%%%%%%%%%%%%%%%%%%%%%%%%%%
\subsection{Inclusion of other carbonaceous grains}
%%%%%%%%%%%%%%%%%%%%%%%%%%%%%%%%%%%%%%%%%%%%%%%%%%%%%%%%%%%%%%%%%%%%%%%

Various populations of carbonaceous grains may constitute cosmic dust 
in addition to graphite.  
We explore the possible significance of other carbonaceous species, 
such as glassy carbon or amorphous carbon.  
We find, however, that the two-component dust model, composed of 
glassy or amorphous carbon and astronomical silicate, does not give 
the extinction curve that lies in the 1 $\sigma$ ranges of observation 
for any $f_{\rm C,glC}/f_{\rm Fe,asil}$ and 
$f_{\rm C,amC}/f_{\rm Fe,asil}$ ratios.  
These carbonaceous grains do not give the proper UV bump at 2175 \AA.  
Glassy carbon shows a broad hump at around 2000 \AA, and amorphous 
carbon has only a broad maximum at around 2000 \AA.  
We consider in the following the model in which graphite is partly
replaced with glassy or amorphous carbon.

Figure 7 shows a few example extinction curves for three-component 
models, compared with our fiducial graphite-silicate model, with 
$q = 3.5$ and $a_{\rm max} =0.24$ $\mu$m, fixed at the same parameters.  
While some of the resulting curves displayed do not satisfy 1 $\sigma$ 
range, as we replace graphite with these components excessively and/or 
we do not tweak the other parameters, we maintain this set of 
parameters to see how the extinction curve is modified with the 
inclusion of other carbonaceous material.

Varying the ($q$, $a_{\rm max}$) parameters, we examine to what
extent graphite can be replaced with glassy or amorphous carbon in the
graphite-silicate model to maintain the extinction curve within the
1 $\sigma$ range.  
Figure 8 presents the maximum fraction of C atoms in glassy 
($f_{\rm C,glC}$) or amorphous carbon ($f_{\rm C,amC}$) relative to the 
entire condensed component of carbon $f_{\rm C,grain}$ against 
$f_{\rm C,grain}/f_{\rm Fe,asil}$, where 
$f_{\rm C,grain} = f_{\rm C,glC} + f_{\rm C,gra}$ or 
$f_{\rm C,grain} = f_{\rm C,amC} + f_{\rm C,gra}$.  
We see that up to $\approx$30--40 \% or $\approx$50--60 \% of carbon 
in grains can be in glassy or amorphous phase, respectively.  
In other words, more than 60--70 \% or 40--50 \% of C atoms must be in 
graphite.  
With this inclusion, the parameters ($q$, $a_{\rm max}$) are shifted 
only a little; see Figure 9.  
Our representative parameters $q=3.5$ and $a_{\rm max} = 0.24$ $\mu$m
still remain to be in the allowed solution.

Silicon carbide (SiC) is another candidate material that can be in
dust grains.  
The upper limit of carbon contained in SiC is approximately 15 \% to 
reproduce the extinction curve (see Figure 8).
This implies that SiC could be a major component of Si-bearing grains.
The abundance of SiC, however, has been tightly limited to $<4$ \% 
of silicon from the lack of 11.3 $\mu$m feature in the extinction 
curve (Whittet et al.\ 1990; Chiar \& Tielens 2006).
The SiC component can be neglected in the discussion of the
extinction curve.

In conclusion, the inclusion of amorphous or glassy carbon has little
effect on the agreement of the graphite-silicate model with 
observations.  The inclusion, however, does not make the overlap of the
two constraints easier.  The dust to gas ratio is unchanged from the
graphite-silicate model.  With SiC the overlap of the two constraints
becomes marginal, and the dust to gas ratio becomes $\rho_{\rm dust}
/\rho_{\rm H} = 1 / (122{+60 \atop -16})$.

%%%%%%%%%%%%%%%%%%%%%%%%%%%%%%%%%%
\begin{figure}
%\epsscale{0.8}
\epsscale{1.1}
\plotone{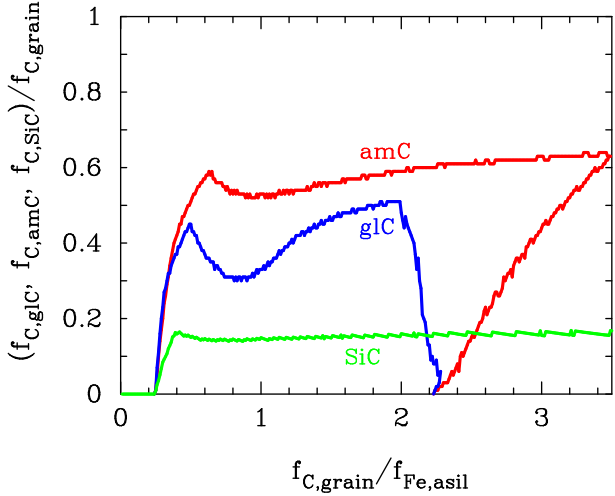}
\caption{ 
Allowed fractions of C atoms in the 
glassy phase ($f_{\rm C,glC}/f_{\rm C,grain}$, blue), 
amorphous phase ($f_{\rm C,amC}/f_{\rm C,grain}$, red), and 
silicon carbide ($f_{\rm C,SiC}/f_{\rm C,grain}$, green).  
The abscissa is $f_{\rm C,grain} / f_{\rm Fe,asil}$, where 
$f_{\rm C,grain} = f_{\rm C,glC} + f_{\rm C,gra}$ or 
$f_{\rm C,grain} = f_{\rm C,amC} + f_{\rm C,gra}$ or
$f_{\rm C,grain} = f_{\rm C,SiC} + f_{\rm C,gra}$.  
The ordinate is the maximally allowed fraction of the individual 
component.
\label{fig8}}
\end{figure}
%%%%%%%%%%%%%%%%%%%%%%%%%%%%%%%%%%

%%%%%%%%%%%%%%%%%%%%%%%%%%%%%%%%%%
\begin{figure}
%\epsscale{0.8}
\epsscale{1.1}
\plotone{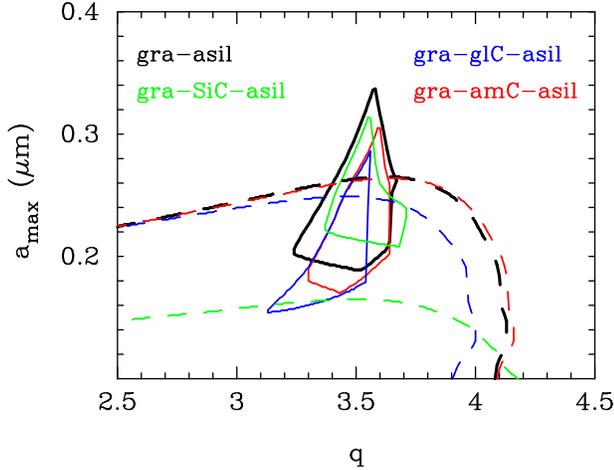}
\caption{ 
Allowed regions of $q$ and $a_{\rm max}$ with which the model 
satisfies the 1 $\sigma$ extinction ranges for the MW when glassy 
carbon, amorphous carbon, or silicon carbide is included in the 
graphite-silicate model.  
The parameters are marginalised over $f_{\rm C,grain}/f_{\rm Fe,asil}$.  
The regions that satisfy the observed size of reddening $E(B-V)$ are 
also shown (dashed curves) for $f_{\rm C,grain} = 0.7$ and 
$f_{\rm Fe,asil} = 1.0$.  
The blue, red, and green lines are, respectively, for the cases where 
glassy carbon, amorphous carbon, and SiC are added to the 
graphite-silicate model, which is represented with the black curves 
for comparison. 
For the $E(B-V)/N_{\rm H}$ constraint it is taken that
$f_{\rm C,glC}/f_{\rm C,grain} = 0.3$, 
$f_{\rm C,amC}/f_{\rm C,grain} = 0.3$, and 
$f_{\rm C,SiC}/f_{\rm C,grain} = 0.1$.
\label{fig9}}
\end{figure}
%%%%%%%%%%%%%%%%%%%%%%%%%%%%%%%%%%

%%%%%%%%%%%%%%%%%%%%%%%%%%%%%%%%%%
\begin{figure}
%\epsscale{0.8}
\epsscale{1.1}
\plotone{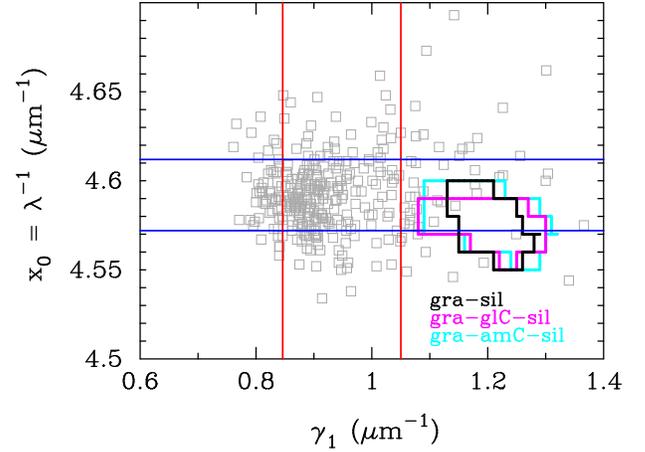}
\caption{
Central wavelength and the width parameter that describe the 2175 
\AA~feature of the MW extinction. 
The observed data from FM07 (squares) are compared to our model with 
the parameters that gives a good fit (within 1 $\sigma$ of the full 
extinction curve) to the global extinction curve. 
The horizontal and vertical lines indicate 1 $\sigma$ of the data.
\label{fig10}}
\end{figure}
%%%%%%%%%%%%%%%%%%%%%%%%%%%%%%%%%%

%%%%%%%%%%%%%%%%%%%%%%%%%%%%%%%%%%%%%%%%%%%%%%%%%%%%%%%%%%%%%%%%%%%%%%%
\subsection{The 2175\AA~Feature}
%%%%%%%%%%%%%%%%%%%%%%%%%%%%%%%%%%%%%%%%%%%%%%%%%%%%%%%%%%%%%%%%%%%%%%%

With small graphite the feature at 2175 \AA~is generated. 
It is known that this feature is observationally fit well with the 
Drude formula, including a smooth background,
\begin{equation}
\frac{A_\lambda}{A_V} = \frac{1}{R_V} \left[ c_0+c_1\lambda^{-1} + 
{c_2\over \gamma_1^2+\lambda^{2} (\lambda^{-2}-\lambda^{-2}_0)^2} 
\right]+1\ .
\end{equation}
In Figure 10 we give the central wavelength $x_0=\lambda_0^{-1}$ and 
the width of the profile $\gamma_1$ for 328 extinction curves of FM07, 
which are summarised as
\begin{equation}
\lambda_0 = 2178 \pm 10 ~{\rm \AA} \hskip10mm
\gamma_1 = 0.95 \pm 0.1 ~\mu{\rm m}^{-1}.
\end{equation}
We fit the feature from our graphite-silicate model, which reads
\begin{equation}
\lambda_0 = 2186 \pm 12 ~{\rm \AA} \hskip10mm 
\gamma_1 = 1.21 \pm 0.08 ~\mu{\rm m}^{-1},
\end{equation}
indicating that the model width is broader by 30 \% 
(see also Draine \& Malhotra 1993), while the central wavelength 
agrees with the observation.  
We also show models with amorphous or glassy carbon included with 
its fraction and size parameters in the range allowed from the entire 
extinction curve.  
Their inclusions make the width slightly ($\approx$10 \%) smaller, 
but not sufficient to give the observed width. 
This seems to be a problem intrinsic to the optical data we adopted 
for graphite.  More detailed treatments may be needed for small 
graphite or PAH.

%%%%%%%%%%%%%%%%%%%%%%%%%%%%%%%%%%
\begin{figure}
%\epsscale{0.6}
\epsscale{1.1}
\plotone{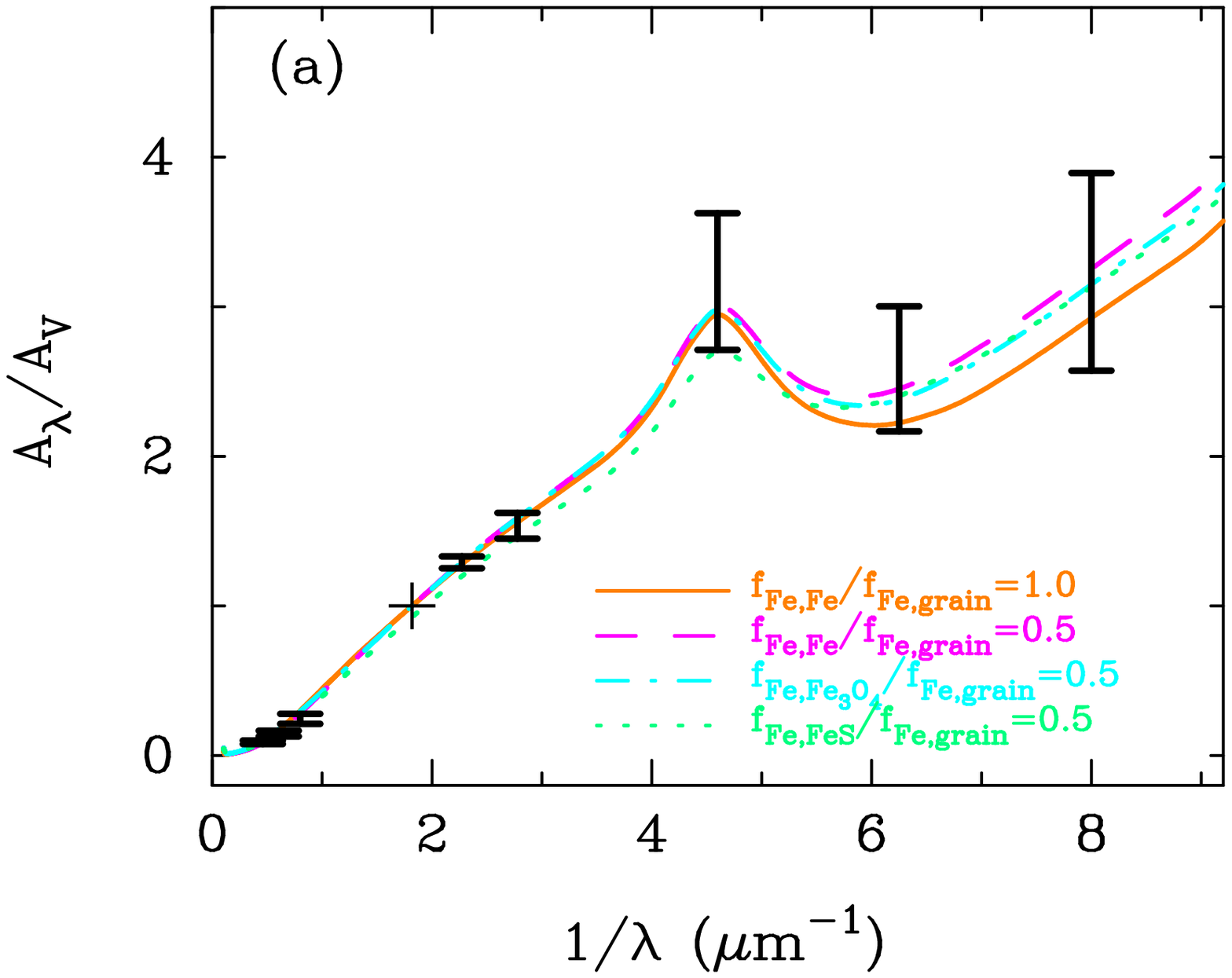}
\vspace{0.8 cm}
\plotone{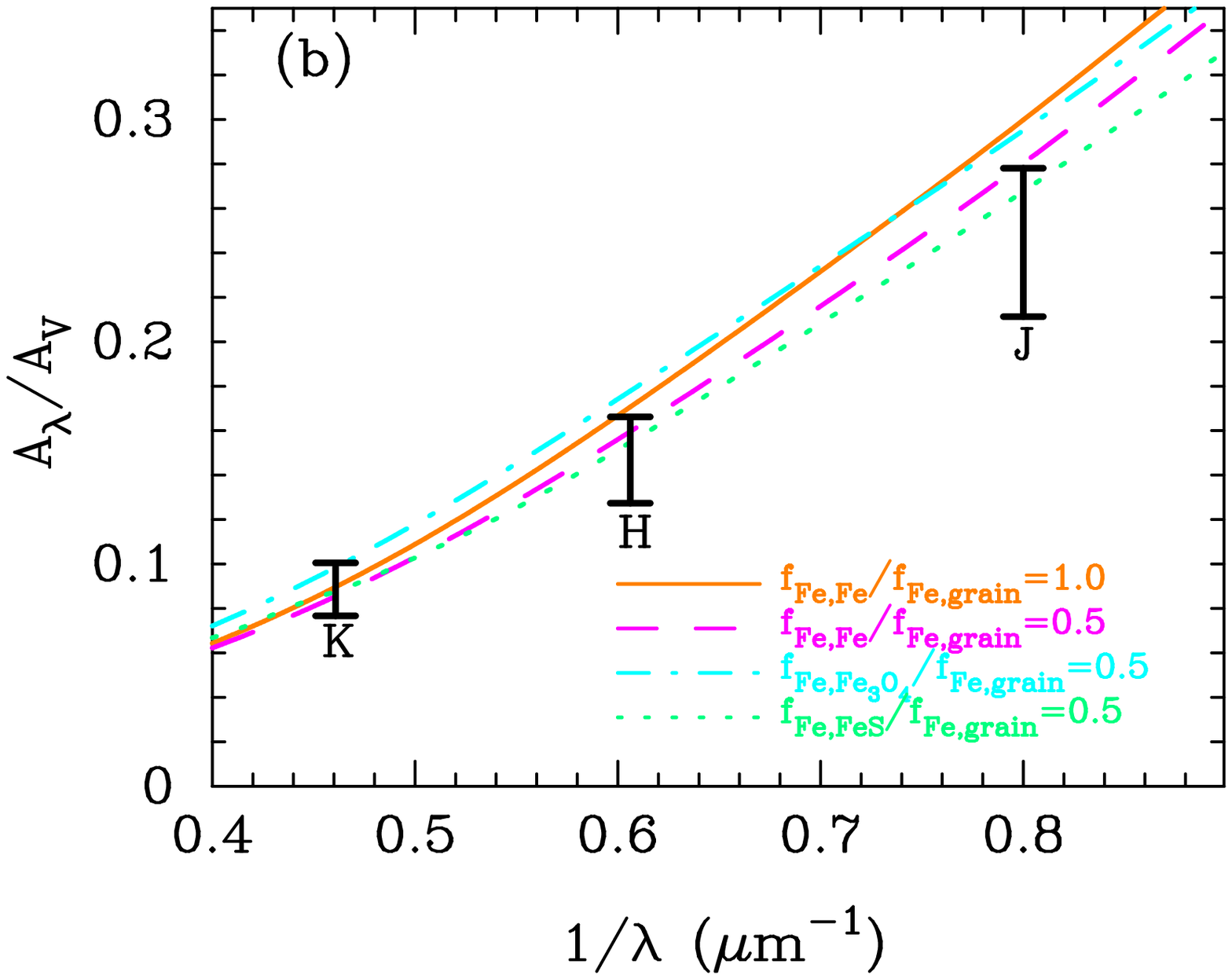}
\caption{ 
(a) Extinction curves for the graphite-silicate model where
iron-bearing silicate is replaced with metallic Fe, Fe$_3$O$_4$ or
FeS, while the corresponding silicate portions are replaced with
forsterite (Mg$_2$SiO$_4$).  
We take the grain-size parameters, $q = 3.5$ and $a_{\rm max} = 0.24$ 
$\mu$m.  
(b) the expanded figure for the NIR.  
We take $f_{\rm C,grain} / f_{\rm Fe,grain} = 0.7$.  
\label{fig11}}
\end{figure}
%%%%%%%%%%%%%%%%%%%%%%%%%%%%%%%%%%

%%%%%%%%%%%%%%%%%%%%%%%%%%%%%%%%%%
\begin{figure}
%\epsscale{0.8}
\epsscale{1.1}
\plotone{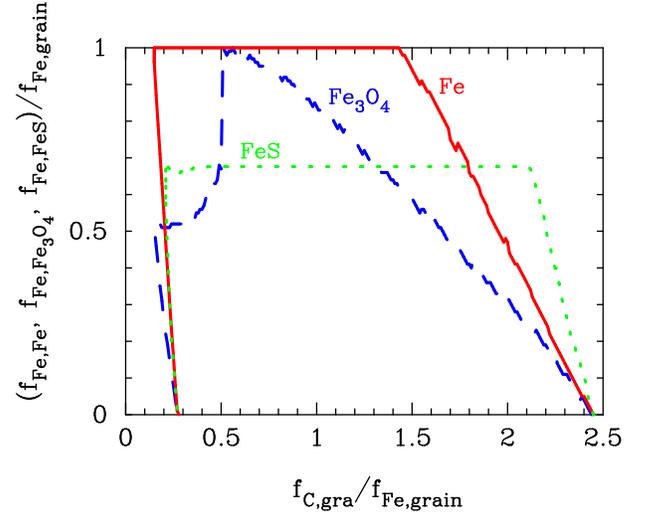}
\caption{
Maximum fractions of Fe atoms allowed in the metallic  
($f_{\rm Fe,Fe}/f_{\rm Fe,grain}$, red solid) or 
the magnetite phase, Fe$_3$O$_4$ 
($f_{{\rm Fe,Fe}_3{\rm O}_4}/f_{\rm Fe,grain}$,  blue dashed), 
with which the model satisfies the 1 $\sigma$ extinction range for the MW.  
The maximum fraction is also drawn for FeS 
($f_{\rm Fe,FeS}/f_{\rm Fe,grain}$, green dotted).  
The abscissa is $f_{\rm C, gra} / f_{\rm Fe, grain}$, where 
$f_{\rm Fe,grain} = f_{\rm Fe,Fe} + f_{\rm Fe,asil}$, 
$f_{\rm Fe,grain} = f_{{\rm Fe,Fe}_3{\rm O}_4} + f_{\rm Fe,asil}$, or 
$f_{\rm Fe,grain} =f_{\rm Fe,FeS}+ f_{\rm Fe,asil}$, and the ordinate 
is the fraction of each species added to the model.
\label{fig12}}
\end{figure}
%%%%%%%%%%%%%%%%%%%%%%%%%%%%%%%%%%

%%%%%%%%%%%%%%%%%%%%%%%%%%%%%%%%%%
\begin{figure}
%\epsscale{0.8}
\epsscale{1.1}
\plotone{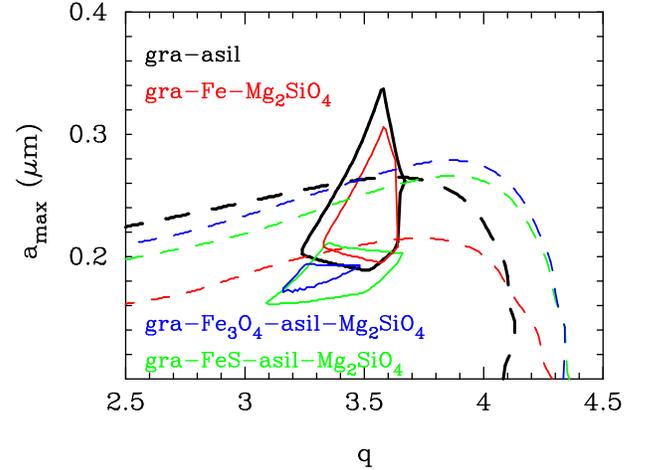}
\caption{ 
Allowed regions of $q$ and $a_{\rm max}$ with which the model 
satisfies the 1 $\sigma$ extinction ranges for the MW, marginalised 
over $f_{\rm C,gra}/f_{\rm Fe,grain}$.  
For the $E(B-V)/N_{\rm H}$ constraint, we take $f_{\rm C,gra} = 0.7$ 
and $f_{\rm Fe,grain}=1.0$.  
The blue, red, and green curves are, respectively, for the inclusion 
of metallic Fe, magnetite 
($f_{{\rm Fe}_3{\rm O}_4}/f_{\rm Fe, grain}=0.8$), and FeS
 ($f_{\rm FeS}/f_{\rm Fe, grain}=0.68$).  
The black curve is our fiducial graphite-silicate model.
\label{fig13}}
\end{figure}
%%%%%%%%%%%%%%%%%%%%%%%%%%%%%%%%%%

%%%%%%%%%%%%%%%%%%%%%%%%%%%%%%%%%%%%%%%%%%%%%%%%%%%%%%%%%%%%%%%%%%%%%%%
\subsection{Inclusion of Fe, Fe$_3$O$_4$, FeS, and Al$_2$O$_3$}
%%%%%%%%%%%%%%%%%%%%%%%%%%%%%%%%%%%%%%%%%%%%%%%%%%%%%%%%%%%%%%%%%%%%%%%

We consider how much iron can be incorporated in generic iron-bearing
grains rather than in astronomical silicate.  
Figure 11 shows the extinction curve for the model with our typical 
parameters $q=3.5$ and $a_{\rm max}=0.24$ $\mu$m.  
Figure 12 exhibits the upper limits of the fraction of Fe in metallic 
phase ($f_{\rm Fe,Fe}$, solid line) and in magnetite, Fe$_3$O$_4$ 
($f_{{\rm Fe,Fe}_3{\rm O}_4}$, dashed line), for the model that 
satisfies the 1 $\sigma$ ranges of the extinction curve.
The rest of Fe atoms in interstellar space are assumed to be condensed
in astronomical silicate, i.e., 
$f_{\rm Fe,grain} = f_{\rm Fe,Fe} + f_{\rm Fe,asil}=1$ or 
$f_{\rm Fe,grain} = f_{{\rm Fe,Fe}_3{\rm O}_4} + f_{\rm Fe,asil}=1$, 
and Mg not included in astronomical silicate is assumed to be in 
forsterite, so that $\sum_{j={\rm grain}} f_{{\rm Mg},j}=1$.  
Here, the abscissa is chosen to be the C/Fe ratio of the condensation.

%%%%%%%%%%%%%%%%%%%%%%%%%%%%%%%%%%
\begin{figure}
%\epsscale{0.6}
\epsscale{1.1}
\plotone{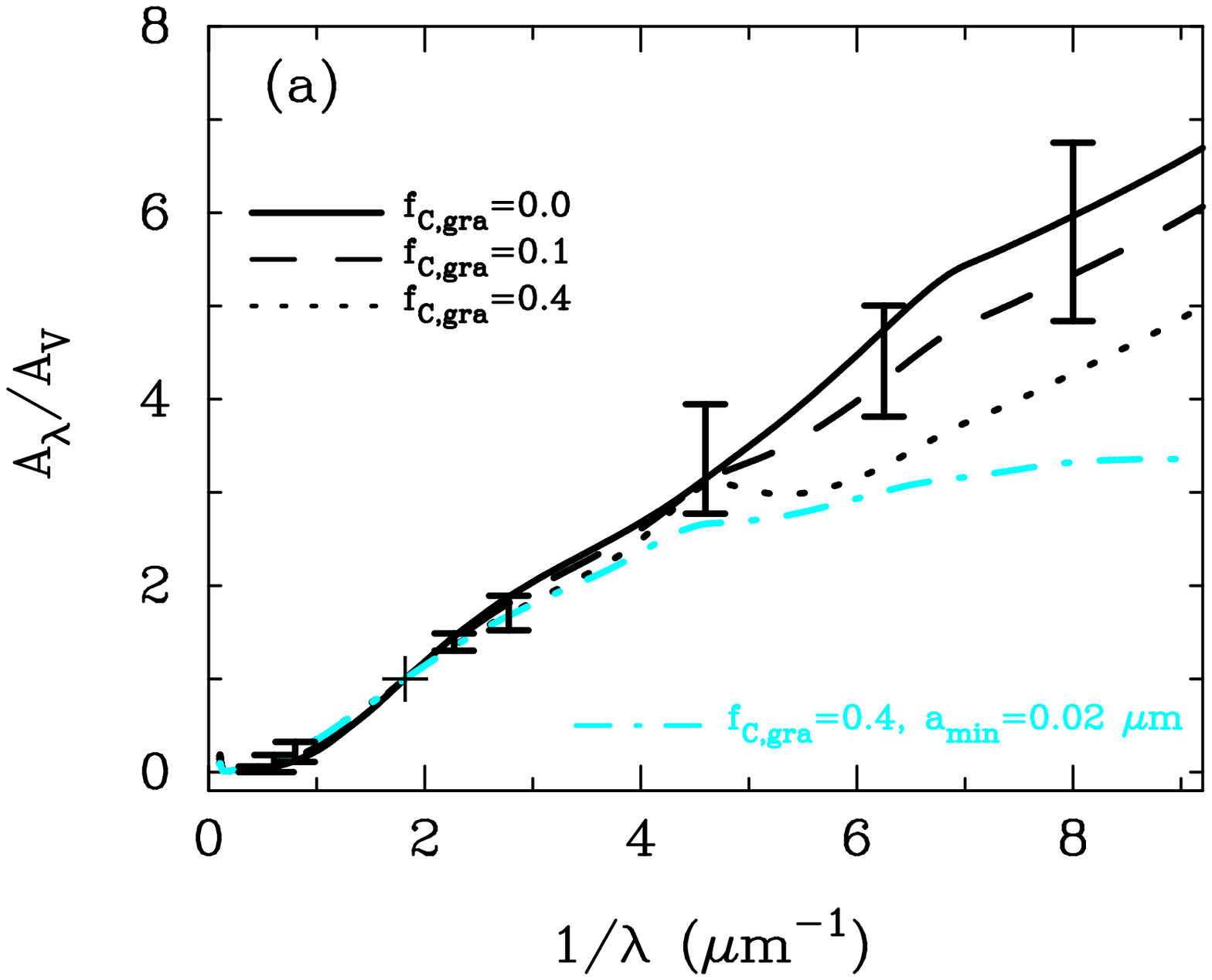}
\vspace{0.8 cm}
\plotone{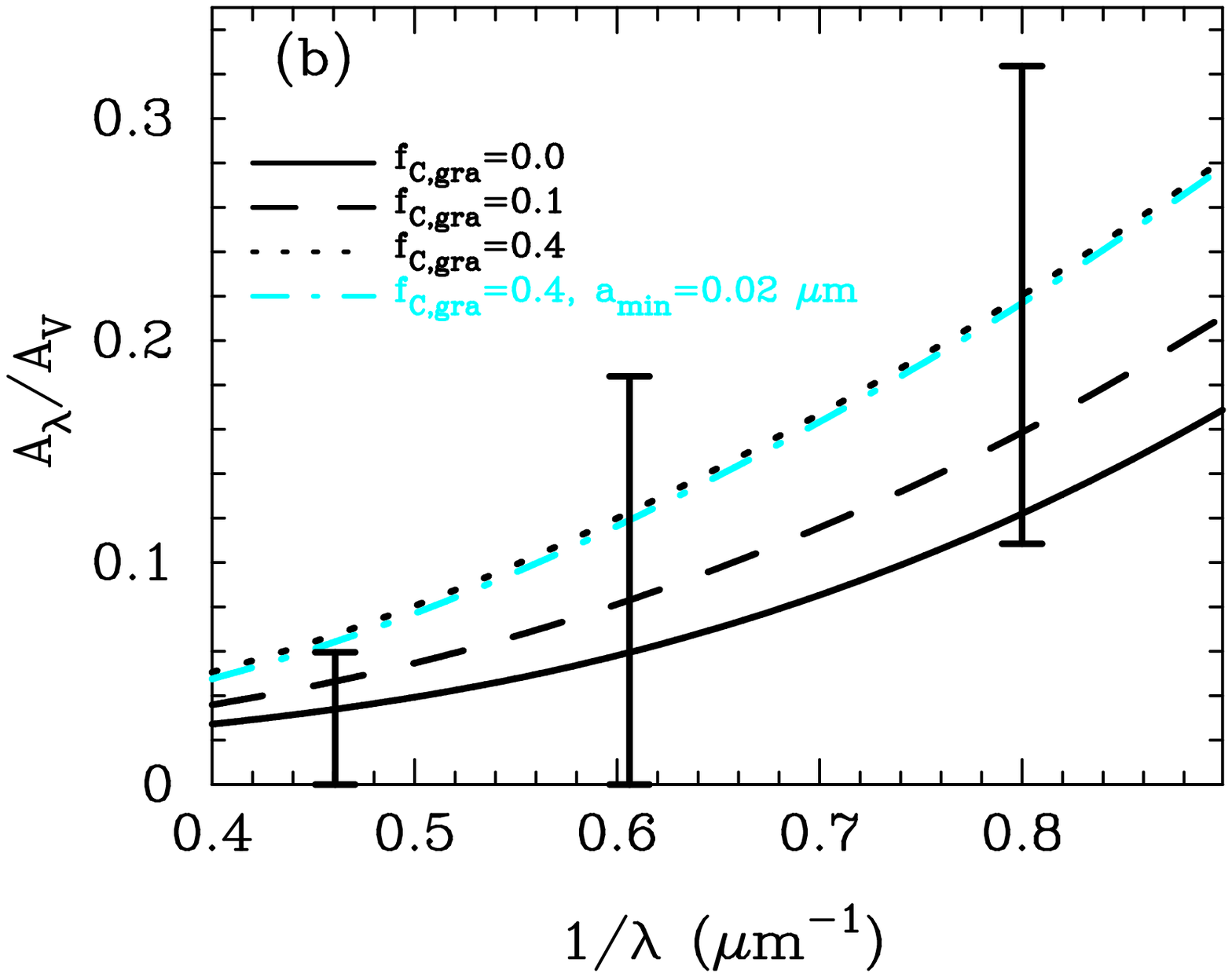}
\caption{
(a) Extinction curves for the SMC from the graphite-silicate dust 
models with $q = 3.5$ and $a_{\rm max} = 0.24$ $\mu$m, and 
(b) the expanded figure of (a) for the NIR.
The solid, dashed, and dotted curves are, respectively, 
for $f_{\rm C,gra} / f_{\rm Fe,asil} =$ 0, 0.1, and 0.4 with 
$a_{\rm min} = 0.005$ $\mu$m fixed as our fiducial choice.
The dot-dashed line is with the minimum cutoff increased to 
$a_{\rm min} = 0.02$ $\mu$m, while 
$f_{\rm C,gra} / f_{\rm Fe,asil} =$ 0.4.
The error bars span the maximum and minimum (taken as \lq 1 $\sigma$')
of the observed SMC extinction curves of G03.
\label{fig14}}
\end{figure}
%%%%%%%%%%%%%%%%%%%%%%%%%%%%%%%%%%

%%%%%%%%%%%%%%%%%%%%%%%%%%%%%%%%%%
\begin{figure}
%\epsscale{0.8}
\epsscale{1.1}
\plotone{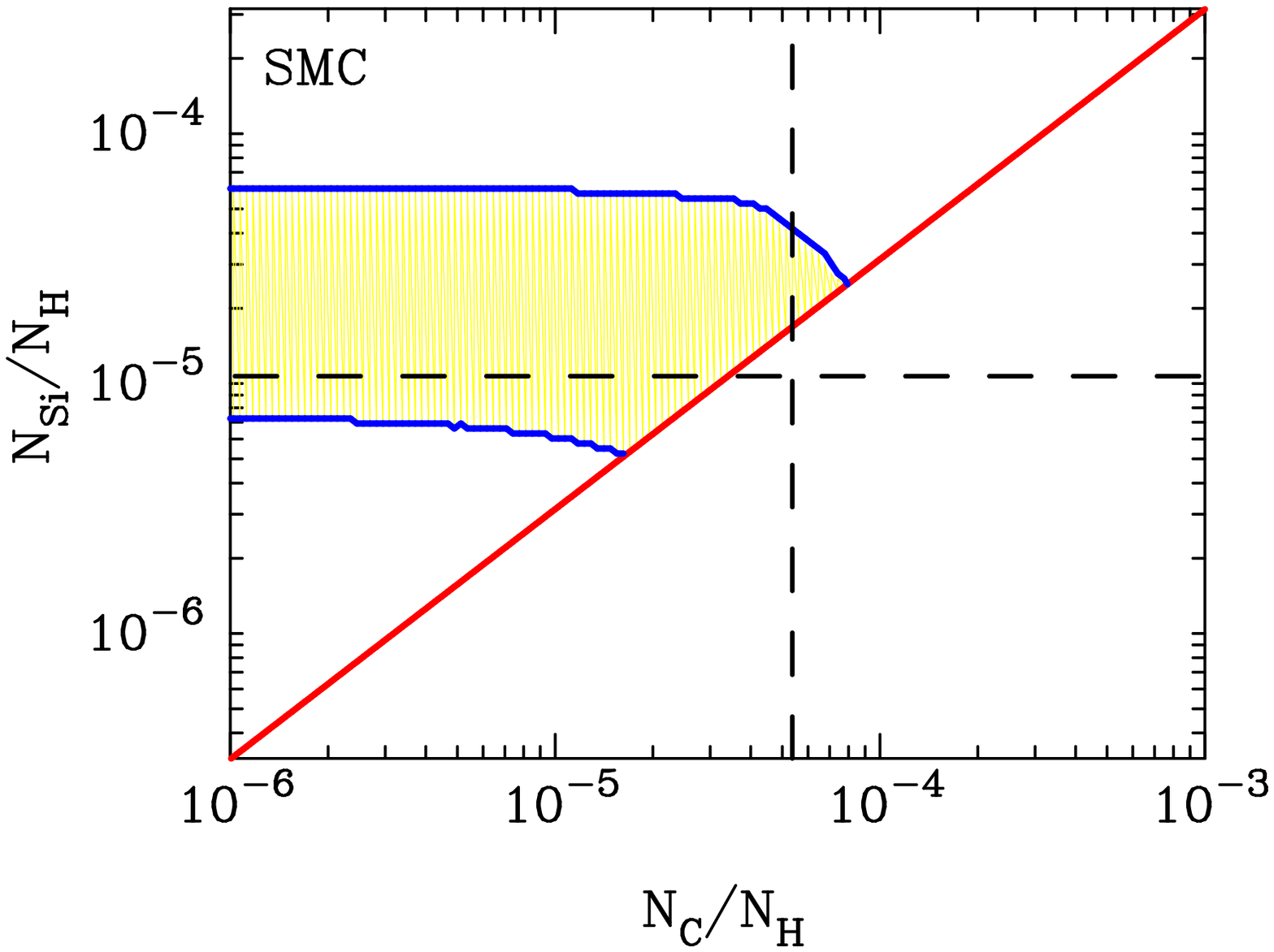}
\caption{ 
The abundance of C and Si in dust grains relative to hydrogen for the 
SMC to meet the extinction curve. 
The range left to the oblique line is allowed from the extinction curve, 
and that within the two curves running nearly horizontal is from 
$E(B-V)/N_{\rm H}$. 
The two dashed lines (horizontal and vertical) show the total abundance 
of C and Si from Russel \& Dopita (1992).
\label{fig15}}
\end{figure}
%%%%%%%%%%%%%%%%%%%%%%%%%%%%%%%%%%

There are parameters ($f_{\rm C, gra}/f_{\rm Fe, grain} =$ 0.15--1.43),
for which Fe may stay entirely in the metallic phase rather than in
astronomical silicate, yet the model gives the correct extinction
curve.  
A similar parameter range, however, is very small with the 
Fe$_3$O$_4$-graphite-forsterite model. 
Practically, the maximum allowed abundance of Fe$_3$O$_4$ is about 
80 \%. 
The NIR extinction of Fe$_3$O$_4$ is large by more than a factor of 2 
compared with the observed extinction, whereas Fe grains give the 
extinction in NIR only by $\approx$30 \% larger, and, therefore, the 
abundance of magnetite is limited from the extinction curve in NIR 
more strongly.  
The 1 $\sigma$-allowed ($q$, $a_{\rm max}$) parameters with these 
models are shown in Figure 13.

%%%%%%%%%%%%%%%%%%%%%%%%%%%%%%%%%%%
\begin{deluxetable*}{lcc}
\tablewidth{0pt}
\tablecaption{Allowed Grain Species}
\tablehead{
\colhead{Dust Grains} &
\colhead{FM07 1 $\sigma$} &
\colhead{CCM 1 $\sigma$}
}
\startdata
& & \\
(1)  Graphite--Astronomical Silicate
      & Yes & No \\
(2)  Glassy Carbon--Astronomical Silicate
      & No & No \\
(3)  Amorphous Carbon--Astronomical Silicate
      & No & No \\
(4)  Graphite--Glassy Carbon--Astronomical Silicate
      & Yes & No \\
(5)  Graphite--Amorphous Carbon--Astronomical Silicate
      & Yes & No \\
(6)  Graphite--SiC--Astronomical Silicate 
      & Yes$^{(a)}$ & No \\
(7)  Graphite-Fe
      & Yes & No \\
(8)  Graphite-Fe$_3$O$_4$
      & No & No \\
(9)  Graphite--Fe--Astronomical Silicate
      & Yes & No \\
(10) Graphite--Fe$_3$O$_4$--Astronomical Silicate
      & Yes & Yes \\
(11) Graphite--Fe--Mg$_2$SiO$_4$
      & Yes & No \\
(12) Graphite--Fe$_3$O$_4$--Mg$_2$SiO$_4$
      & Yes & Yes \\
(13) Graphite--Fe--Astronomical Silicate--Mg$_2$SiO$_4$
      & Yes & No \\
(14) Graphite--Fe$_3$O$_4$--Astronomical Silicate--Mg$_2$SiO$_4$
      & Yes & Yes \\
(15) Graphite--FeS--Astronomical Silicate--Mg$_2$SiO$_4$
      & Yes & No \\
\enddata
\tablecomments{
We count the species when its fraction is more than 10 \%.
(a): this model is ruled out if we consider the lack of 11.3 
$\mu$m feature.}
\end{deluxetable*}
%%%%%%%%%%%%%%%%%%%%%%%%%%%%%%%%%%

The combination of metallic Fe and forsterite works in a way virtually
the same as astronomical silicate, which means that the Mg/Fe ratio in 
astronomical silicate is arbitrary.  
The situation is somewhat different if iron is in magnetite.  
The inclusion of Fe$_3$O$_4$ disturbs the allowed range in 
($q$, $a_ {\rm max}$) plane by an appreciable amount, making 
$a_ {\rm max}$ smaller by 20 \%.
Nevertheless, we see that the overlap of the two constraints is
maintained.

The cosmic abundance of sulphur is about 0.7 times that of Fe.  
We consider the case where all S is in troilite, FeS. 
The rest of Fe is in astronomical silicate and further the rest of Mg 
is in forsterite.
Troilite also gives large NIR extinction almost as much as
magnetite, but maximal amount of FeS is allowed because of smaller
cosmic abundance of sulphur.  
This case satisfies the 1 $\sigma$ constraint with $a_{\rm max}$ 
somewhat smaller than the graphite-silicate model, as seen in Figure 13.

In conclusion iron can be in a variety of grain species, including
olivine, metallic phase, Fe$_3$O$_4$, or FeS without disturbing the
extinction curve.  
The only condition is that Fe is not predominantly in magnetite, 
which produces too large NIR extinction.
The grain size parameters are nearly the same for all cases, up to the
result that the maximum size cutoff becomes by 20--30 \% smaller if
Fe$_3$O$_4$ or FeS is the major component. 
All iron atoms need not necessarily be locked in astronomical silicate, 
and the ratio of Fe:Mg in astronomical silicate is arbitrary.  
Allowing for the inclusion of a variety of iron material, the mass 
density of dust differs little from the graphite-silicate model.

We can ignore the contribution from corundum (Al$_2$O$_3$).  
In addition to the small abundance of Al (1/13 of Mg), the $Q$ factor 
for corundum is small (Toon et al.\ 1976).  
Even if all Al atoms are locked in corundum, they contribute little to 
modifying the extinction curve.

Table 4 summarises our result for most grain species we considered in
the present work.  
Limiting ourselves to at most four grain species, we consider what 
combination of species would give the extinction curve consistent with 
the observation at 1 $\sigma$.  
Some of them were already discussed in the text above.  
We take the species as valid when its fraction is more than 10 \%.  
We here note one particular case that can reproduce the extinction law 
of CCM: 
it is the combination of graphite, astronomical silicate, and magnetite.
This is due to the particularly large $Q$ factor of magnetite in the
NIR, which is excluded if we take FM07.

%%%%%%%%%%%%%%%%%%%%%%%%%%%%%%%%%%
\begin{figure}
%\epsscale{0.8}
\epsscale{1.1}
\plotone{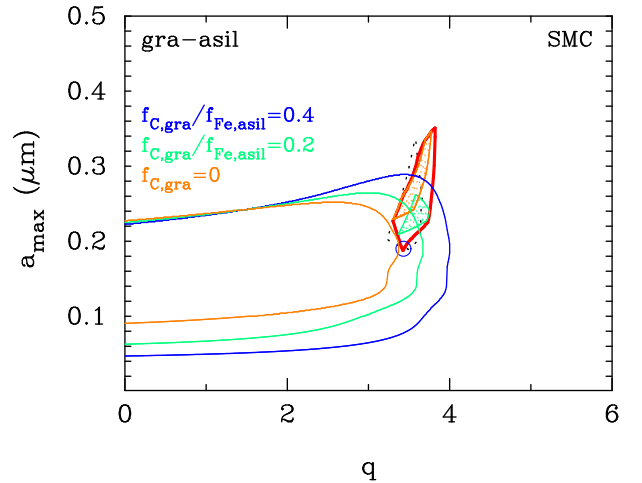}
\caption{ 
Allowed regions of $q$ and $a_{\rm max}$ with which the model satisfies 
the \lq 1 $\sigma$' extinction ranges for the SMC.  
The graphite-to-silicate ratio is marginalised over 
$0 \le f_{\rm C,gra} / f_{\rm Fe,asil} \le 0.41$.  
The inlaid curves indicate the regions for $f_{\rm C,gra} = 0$, 0.2,
and 0.4.  
The regions for 0.4 is a point indicated with a small circle (the 
actual region is enlarged) near the bottom of the allowed region.  
The constraints from the reddening $E(B-V)/N_{\rm H}$ are shown with 
brown, green, and blue contours for $f_{\rm C,gra} = 0$, 0.2, and 0.4 
($f_{\rm Fe,asil} = 1.0$).  
The thin dotted curve recapitulates the allowed region for MW dust in 
Figure 6 for comparison.
\label{fig16}}
\end{figure}
%%%%%%%%%%%%%%%%%%%%%%%%%%%%%%%%%%

%%%%%%%%%%%%%%%%%%%%%%%%%%%%%%%%%%%%%%%%%%%%%%%%%%%%%%%%%%%%%%%%%%%%%%%
\section{Dust in the SMC}
%%%%%%%%%%%%%%%%%%%%%%%%%%%%%%%%%%%%%%%%%%%%%%%%%%%%%%%%%%%%%%%%%%%%%%%

Figure 14 shows examples of the model extinction curve for the SMC
with $q=3.5$ and $a_{\rm max}=0.24$ $\mu$m for 
$f_{\rm C, gra}/f_{\rm Fe,asil} =$ 0, 0.1, and 0.4.  
(The curve for $f_{\rm C,gra}/f_{\rm Fe,asil} =0.4$ appears to be out 
of the 1 $\sigma$ range, but it satisfies 1$\sigma$ 
if $a_{\rm max}\simeq 0.19$ $\mu$m.)  
The model for $f_{\rm C,gra}/f_{\rm Fe,asil} \le 0.4$ reproduces the 
SMC extinction curve within 1 $\sigma$ when $q$ and $a_{\rm max}$ are 
adjusted.  
With $f_{\rm C,gra}/f_{\rm Fe,asil} = 0.4$ one sees in this figure the
symptom that a small 2175 \AA~bump starts appearing and the rise
towards the far UV side becomes insufficient.  
We draw another curve (dot-dashed curve) with the minimum grain size 
cutoff increased to $a_{\rm min}=0.02$ $\mu$m, removing much of small 
graphite grains. 
This curve lies similar to the one suggested by Calzetti et al.\ 
(1994) for star-forming galaxies.
The bump at 2175 \AA~is reduced and becomes insignificant, but the
removal of small size grains makes the far UV rise insufficient.
Therefore, such a case is excluded.  
Within the graphite-silicate model, the way to make the predicted 
extinction consistent with the observation is not to change the grain 
size parameters but to decrease the abundance of graphite grains. 
The steeper rise towards far UV is accounted for by the  
smaller silicate grains. 

The abundance of Si and C contained in dust is shown in Figure 15,
where the abundance from Russell \& Dopita (1992) is also indicated
with dashed lines.  
We have $\log ({\rm Si/H}) + 12 =$ 6.7--7.8 and 
$\log({\rm C/H}) + 12 \leq 7.9$ from the extinction in the SMC. 
When $f_{\rm Si,grain} = 1$, $\log ({\rm C/H}) + 12 \leq 7.6$ in 
agreement with eq. (\ref{eq:smcCgra}) below. 
It is noted that the carbon abundance in graphite in the SMC is at 
least a factor of 1.5 times smaller than the abundance of Russell \& 
Dopita (if Si is all condensed into dust; following the horizontal 
dashed line); the latter gives 7.73 for C and 7.03 for Si.
The upper and lower curves correspond to the 1 $\sigma$ error of
$N_{\rm H}/E(B-V)$. 
For the SMC $\rho_{\rm gra} / \rho_{\rm asil} = 
0.54 f_{\rm C,gra} / f_{\rm Fe,asil} \leq 0.22$.

Figure 16 is a summary of the region of $q$ and $a_{\rm max}$, where
the graphite-silicate model gives the extinction curve within
the 1 $\sigma$ range marginalised over $f_{\rm
  C,gra}/f_{\rm Fe,asil}$.  The allowed grain size distribution
lies again for $3.3
\la q \la 3.8$ and 0.19 $\mu$m $\la a_{\rm max} \la 0.35$ $\mu$m.  It
is interesting to observe that this range for the SMC agrees with that
for the MW (recapitulated in the figure with the dotted curve) in spite
of the significantly different behaviour of the two extinction curves,
as was noted earlier in Pei (1992).

The other curves in Figure 16 show the regions allowed by 
$N_{\rm H}/E(B-V)$ for the 1 $\sigma$ error of Eq (2).  
The overlap of the two constraints is seen for
\begin{equation}
f_{\rm C,gra} < 0.41.
\label{eq:smcCgra}
\end{equation}
This allowed maximum fraction of graphite is smaller than the 
minimum $f_{\rm C,gra}/f_{\rm Fe,asil}$ favoured for the MW
($f_{\rm C,gra} \ge 0.56$), suggesting that condensation of carbon
into grains should be less efficient in the SMC, as might be satisfied 
if, e.g., $f_{\rm C,gra} \propto N_{\rm C}$. 
More detailed arguments, however, depend on the accuracy of the 
abundance estimates, in particular, for the SMC.

For the model that satisfies 1 $\sigma$ constraints of both extinction
curve and $E(B-V)$ size for the SMC, we find 
$\rho_{\rm dust} / \rho_{\rm H} = 1 /{760 {+150\atop -660}}$,
where the large error range comes from that of $E(B-V)/N_{\rm H}$, and
corresponds roughly to the upper and lower curves of the allowed
region in Figure 15.  
If the abundance is constrained with Russell \& Dopita's value and Fe 
is all condensed,
\begin{equation}
\rho_{\rm dust} / \rho_{\rm H} = 1 /{760 {+70 \atop -90}}.
\end{equation}

We have
\begin{equation}
K_{\rm ext} = (2.2 \pm 0.3)\times 10^4 ~~{\rm mag ~cm^2 ~g^{-1}}.
\end{equation}
(It is $(2.1 \pm 0.4) \times 10^4$ mag cm$^2$ g$^{-1}$ if the 
abundance constraints are removed.)
The ratio of graphite to silicate is 
$\rho_{\rm gra} / \rho_{\rm asil} = 0.11 \pm 0.11$, or 
$N_{\rm C} / N_{\rm Fe} = 1.6 \pm 1.6,$ in numbers of atoms.  
This ratio is 4.5 times smaller than that for the MW.  
The abundance itself indicates the ratio of C/Fe to be 1.4 times
smaller, so the formation of carbonaceous grains is suppressed by
$\approx$3 times more in the SMC.

%%%%%%%%%%%%%%%%%%%%%%%%%%%%%%%%%%%%%%%%%%%%%%%%%%%%%%%%%%%%%%%%%%%%%%%
\section{Conclusions and discussion}
%%%%%%%%%%%%%%%%%%%%%%%%%%%%%%%%%%%%%%%%%%%%%%%%%%%%%%%%%%%%%%%%%%%%%%%

We have confirmed that the graphite-silicate grain model gives a 
satisfactory description of the extinction curves within the 
{\it simple power-law model} of the size distribution.  
The grain size distribution is tightly constrained to the index 
$q = 3.5 \pm 0.2$.
We showed that departures from a power law are not needed. 
However, we need to cutoff the power-law at some 
maximum size, $a =$ 0.2--0.3 $\mu$m.  
Grains may be a variety of carbonaceous and silicate 
materials. 
Their size parameters, however, vary little and the same parameters
also apply to both MW and SMC extinction curves in spite of their 
apparently different behaviour.  
The difference between the two extinction curves can be ascribed to 
the abundance of graphite relative to silicates, and hence to some 
lower efficiency ($\lesssim 1/2$) of graphite condensation, beyond 
the lower ratio ($\lesssim 1/1.5$) of carbon to silicon in the 
SMC indicated in abundance estimates currently available.
We also remarked on the somewhat dissatisfying description of the 2175 
\AA~feature.

While grains can be a variety of combinations that contain silicate 
and carbonaceous material, the presence of a significant abundance of
graphite is important.  
It is interesting to note that the resulting $q = 3.5$ is the power 
expected from collisional equilibrium for small grains. 
We have derived the elemental abundance of Si and C contained in grains 
from the extinction.  For the MW
it is consistent with the solar.  
If we take the widely adopted abundance of GS98, we infer the fraction 
of carbonaceous grains against total carbon, 
$f_{\rm C,gra} =$ 0.6--0.7, which agrees with the observational 
depression factor for carbon. 
This may be compared with the corresponding value for the SMC, 
$f_{\rm C,gra} \lesssim 0.4$, taking the elemental abundance of 
Russell \& Dopita (1992). 
The required silicate is also consistent.

The upper cutoff is compelling for the grain size distribution to give 
the correct shape of the extinction curve.
Extending it to a larger size would disrupt the agreement with
observations (see Draine 2009): 
for example, the $R_V$ parameter becomes intolerably 
large.\footnote{This is caused by the fact that the $R_V$ parameter 
  resulting from silicate grains is wildly oscillating as a function of 
  $a$ for $a > 0.2$ $\mu$m, which makes the resulting $R_V$ parameter 
  too large.}  
While we are not able to find the reason for the cutoff, we may see 
another argument that forces us to impose the presence of a cutoff in 
the power law.  
The abundance of grains is constrained at a submicron region by
the amount of the observed optical extinction.  
The power law with $q = 3.5$ means that the integrated mass of grains 
is slowly increasing with the upper cutoff mass $m_c$ as 
$m_c^{0.17}$.  
If the cutoff were larger, we would have too large a mass in grains to 
be accounted for as a product of stellar evolution.  
The dust mass density relative to hydrogen 
$\rho_{\rm dust} / \rho_{\rm H} \approx 0.008$ means 
$\Omega_{\rm dust} \approx 4.4 \times 10^{-6}$ taking the
global hydrogen abundance for HI and H$_2$ observations, 
$\Omega_{\rm H} = 5.4 \times 10^{-4}$.  
If dust is a product of stellar evolution over the cosmic time, stars 
of the amount $\Omega_{\rm star}\simeq 0.003$ would produce dust grains 
no more than $\Omega_{\rm dust}\approx 1\times 10^{-5}$ (Fukugita 2011).  
This amount is consistent with the power-law distribution constrained 
from the extinction if the power-law is cutoff at 0.25 $\mu$m.  
If the maximum size cutoff were an order of magnitude larger, say, 
the dust abundance 
would be larger by $\approx$50\%, more than the star formation 
activity can account for.

Small astronomical objects may also obey the same power law.  
Their normalisation, however, should be smaller by a large factor than 
that for the dust grains.  
The integrated mass density of the core of planets for $a > 100$ km is
estimated to be 1/300 that of dust grains (Fukugita 2011).  
The power law should be broken by two orders of magnitude at a few tenths of 
micron.
The addition of the planet core mass disturbs little the estimate of
the mass density borne by small objects.

We have emphasized that the behaviour of the extinction curve in NIR is
important. 
If the power of the NIR extinction curve $\lambda^{-\gamma}$ is as small 
as $\gamma=1.6$, as was derived earlier and adopted by CCM or WD01, one 
cannot reproduce the extinction curve from UV to NIR by the grain model 
with a simple power law of the grain size.  
One needs to substantially adjust the size distribution,
by, e.g., adding extra components,   
as was done by WD01, Zubko 
et al.\ (2004), and others.  
With a larger power of the NIR wavelength dependence, such as 
$\gamma=$ 1.8--2.1, however, the simple power law model works 
for the observed extinction.  
The NIR power $\gamma=1.6$ can be consistent with the power law grain 
size only when the iron component of grains is largely in 
magnetite, which has a large NIR scattering efficiency.

We have seen that the presence of graphite grains is uniquely
important, at least in the MW, although the condensation of C atoms 
may not necessarily be all into graphite.  
Roughly half the amount of carbon may be condensed into a glassy or an 
amorphous phase,
whereas some significant fraction of C atoms must be in graphite.  
Iron may also remain in the metallic phase.  
There is no need for all Fe atoms to be locked in astronomical silicate.  
The effective ratio of Mg to Fe in olivine seems arbitrary.  
Fe$_3$O$_4$, however, cannot be predominant.  
Fe may also be in troilite, as much as the sulphur abundance allows.

The size distribution is well converged to a narrow range, regardless
of whether other grain compositions are included.  The variation of
extinction along lines of sight may be accounted for by a small
variation of the grain-size distribution with $\Delta q \approx 0.2$
and/or $\Delta a_{\rm max}/a_{\rm max} \approx 0.3$, and with changing the
graphite to silicate ratio as for the difference between the MW and the
SMC.  It is noteworthy that the $a^{-3.5}$ power as expected in
collisional equilibrium seems to be generic to dust grains.

%%%%%%%%%%%%%%%%%%%%%%%%%%%%%%%%%%%%%%%%%%%%%%%%%%%%%%%%%%%%%%%%%%%%%%%
\acknowledgments
%%%%%%%%%%%%%%%%%%%%%%%%%%%%%%%%%%%%%%%%%%%%%%%%%%%%%%%%%%%%%%%%%%%%%%%

We would like to thank Bruce Draine for many discussions and 
carefully reading the manuscript, which have improved our work 
significantly. We also thank Kevin Bundy for many suggestions
on the manuscript.
The work is supported in part by the Grants-in-Aid for Scientific 
Research of the Japan Society for the Promotion of Science 
(22684004, 23224004, 23540288).  
MF is supported by the Monell foundation and the W.M. Keck foundation
at Princeton.

%\newpage
%%%%%%%%%%%%%%%%%%%%%%%%%%%%%%%%%%%%%%%%%%%%%%%%%%%%%%%%%%%%%%%%%%%%%%%

%%%%%%%%%%%%%%%%%%%%%%%%%%%%%%%%%%%%%%%%%%%%%%%%%%%%%%%%%%%%%%%%%%

\end{document}